\documentclass[superscriptaddress,
twocolumn,secnumarabic,nofootinbib]{revtex4-1}
\usepackage{amsmath}
\usepackage[utf8]{inputenc}
\usepackage[T1]{fontenc}

\pdfoutput=1

\usepackage{bm}
\usepackage{times}
\usepackage{braket}
\usepackage{color,graphicx}
\usepackage[utf8]{inputenc}
\usepackage[T1]{fontenc}
\usepackage[export]{adjustbox}

\usepackage{enumitem}

\usepackage{verbatim}

\usepackage{float}

\usepackage{colortbl}
\usepackage{longtable}
\usepackage{tabularx}
\usepackage{makecell}

\usepackage{multirow}

\usepackage{xcolor}

\usepackage{lipsum}

\usepackage[normalem]{ulem}

\definecolor{LightGray}{gray}{0.96}
\definecolor{Gray}{gray}{0.94}
\definecolor{nicered}{rgb}{0.7,0.1,0.1}
\definecolor{nicegreen}{rgb}{0.1,0.5,0.1}

\usepackage{hyperref}
\hypersetup{colorlinks,citecolor=nicegreen,
linkcolor=nicered,urlcolor=black}
\hypersetup{colorlinks=true}

\allowdisplaybreaks

\setlongtables

\begin{document}

\title{New physics in $b\to s\ell\ell$ transitions at one loop}

\author{Rupert Coy} \email[]{rupert.coy@umontpellier.fr}
\author{Michele Frigerio} \email[]{michele.frigerio@umontpellier.fr}

\affiliation{Laboratoire Charles Coulomb (L2C), University of Montpellier, CNRS, Montpellier, France}

\author{Federico Mescia}  \email[]{mescia@ub.edu}

\affiliation{Departament de Física Quàntica i Astrofísica (FQA), Institut de Ciències del Cosmos
(ICCUB), Universitat de Barcelona (UB), Spain}

\author{Olcyr Sumensari}   \email[]{olcyr.sumensari@pd.infn.it}
\affiliation{Dipartimento di Fisica e Astronomia ``G.\ Galilei'', Universit\` a di Padova, Italy \\
Istituto Nazionale Fisica Nucleare, Sezione di Padova, I-35131 Padova, Italy}  
  
\begin{abstract}
We investigate new-physics contributions to $b\to s \ell\ell$ transitions in the context of an effective field theory extension of the Standard Model, including operator mixing at one loop.
We identify the few scenarios where a single Wilson coefficient, $C/\Lambda^2 \sim 1/{\rm TeV}^2$, induces a substantial shift in the lepton flavour universality ratios $R_K$ and $R_{K^*}$ at one loop, while evading $Z$-pole precision tests, collider bounds, and other flavour constraints. Novel fits to the present data are achieved by a left-handed current operator with quark-flavour indices $(2,2)$ or $(3,3)$. 
Interestingly, the running of the Standard Model Yukawa matrices gives the dominant effect for these scenarios. 
We match the favoured effective-theory scenarios to minimal, single-mediator models, which are subject to additional stringent constraints. 
Notably, we recognise three viable instances of a leptoquark with one coupling to fermions only. 
If the anomalies were confirmed, it appears that  
one-loop explanations have good prospects of being directly tested at the LHC.

\end{abstract}
\maketitle

\section{Introduction}
\label{sec:intro}

For the past several years, semileptonic $B$-meson decays have exhibited an intriguing pattern of deviations from the Standard Model (SM) predictions. Data indicate an apparent violation of lepton flavour universality (LFU), that is, $B$-mesons decay with different rates into different lepton flavours. 
The most compelling observations come from LHCb measurements of the theoretically clean~\cite{Aaij:2017vbb,Aaij:2019wad}
\begin{equation}
R_{K^{(\ast)}}^{[q_1^2,q_2^2]} 
= \dfrac{\mathcal{B}^\prime(B\to K^{(\ast)}\mu^+\mu^-)}{\mathcal{B}^\prime(B\to K^{(\ast)}e^+e^-)} \,,
\end{equation}
where $\mathcal{B}^\prime$ stands for the partial branching fraction integrated in the interval $q^2\in[q_1^2,q_2^2]$ of dilepton squared-momenta.
The reported values of $R_{K^{(\ast)}}$ in different $q^2$-bins are consistently smaller than the SM predictions~\cite{Hiller:2003js},
providing motivation for new-physics contributions to $b\to s\ell\ell$ transitions. 
A further departure from LFU has been observed in exclusive $B$-meson decays based on $b\to c\ell\nu$ transitions ($\ell=e,\mu,\tau$) \cite{Lees:2012xj,Huschle:2015rga,Aaij:2015yra,Hirose:2016wfn}, which may also point to physics beyond the SM.

Since no clear evidence of new physics has been found in direct searches at the LHC, it is reasonable to assume that new degrees of freedom have masses well above the electroweak scale. In this case, an effective field theory (EFT) respecting the full SM gauge symmetry, known as the SMEFT, provides the most appropriate description of data~\cite{Buchmuller:1985jz,Grzadkowski:2010es}. Within this framework, the $b\to s\ell\ell$ and $b\to c\ell \nu$ anomalies point to very different scales of new physics ~\cite{DiLuzio:2017chi}, 
namely $M/g_{\mathrm{NP}} \sim 20$~TeV and 2~TeV respectively, where $g_{\mathrm{NP}}$ denotes a generic tree-level coupling between the SM fermions and new states of mass $M$. 
Given the present exclusion limits from direct searches and assuming perturbative couplings, the charged-current anomalies can only be explained via tree-level contributions, while the neutral-current ones can potentially be explained by tree or loop-level contributions.

In this paper, we systematically determine which scenarios can significantly contribute to $R_{K^{(\ast)}}$ at loop level. 
While tree-level contributions require states with mass $M \sim 20\ {\rm TeV} \times g_{\mathrm{NP}}$, in the case of operator mixing at one loop we obtain, instead, $M \sim 20\ {\rm TeV} \times g_{\mathrm{NP}} \times (g_{\mathrm{SM}}/4\pi)$, which brings the new physics scale close to the one currently probed by direct searches at the LHC.
Tree-level EFT contributions to $R_{K^{(\ast)}}$ within the SMEFT were first identified in Ref.~\cite{Alonso:2014csa} and quantitatively studied in e.g.~Refs.~\cite{Celis:2017doq,Aebischer:2019mlg,Ciuchini:2019usw}. 
One-loop solutions have been less extensively studied, despite being the most intriguing option for phenomenology. 
We aim to address two main questions. 
Is there room for new physics close to the TeV scale, despite the existing direct searches, and electroweak and flavour constraints? If room is left, which light states are expected and how can they be tested at the LHC? 
Some one-loop contributions to $b\to s\ell\ell$ have already been identified in Ref.~\cite{Celis:2017doq}. In this article, we will perform a more comprehensive analysis, considering all possible Wilson coefficients (WCs) and flavour indices 
within a complete basis of dimension-six SMEFT operators, and using the latest experimental results.

The loop effects can be computed using the renormalisation group equations (RGEs) of operators introduced at some new physics scale $\Lambda$, which is assumed to be larger than the electroweak scale~\cite{Jenkins:2013zja,Jenkins:2013wua,Alonso:2013hga}. 
Operator mixing is also important for identifying complementary experimental constraints on a given WC, see e.g.~Ref.~\cite{Feruglio:2016gvd}. 
We will consistently take into account all relevant one-loop mixing effects to assess the viability of each scenario, studying an extended collection of experimental constraints with respect to previous analyses. 
Finally, we will build single-mediator simplified models, which provide an explicit realisation of the viable EFT scenarios,
and we will account for additional, model-dependent bounds on the relevant mediators. 
Our general classification of new physics contributions to $R_{K^{(\ast)}}$ will be independent of the current experimental values, which are not yet settled, hence our analysis will remain pertinent when the time comes to reinterpret updated experimental results.

The remainder of this paper is organised as follows. In Section \ref{sec:efts}, we introduce the effective Lagrangian describing the $b\to s\ell\ell$ transition at tree-level and confront it with the $R_{K^{(\ast)}}$ anomalies. In Sec.~\ref{sec:loop-level}, we extend our discussion to loop-level contributions via an analysis of the RGEs. The viable loop-level EFT scenarios are characterised in detail in Sec.~\ref{sec:viable}, and the simplified models matching onto these scenarios are presented in Sec.~\ref{sec:simplified}. 
Our findings are summarised in Sec.~\ref{sec:conclusion}.

\section{Effective theory for semi-leptonic decays}
\label{sec:efts}

\subsection{Low-energy weak effective description}
\label{ssec:lowenergy}

The effective Lagrangian used to describe $b\to s\ell_i\ell_i$ transitions can be written as

{
\begin{align}
\label{eq:leff-mb}
\mathcal{L}_{\mathrm{WET}}= \dfrac{4 G_F \lambda_t}{\sqrt{2}} \sum_{i,a} 
&C_a^{ii}(\mu)\, \mathcal{O}_a^{ii}(\mu)
+ \mathrm{h.c.}\,,
\end{align}
}

\noindent where $\lambda_t=V_{tb}\, V_{ts}^\ast$, and $C_{a}^{ii}$ denote the relevant Wilson coefficients, which should be evaluated at $\mu=m_b$. For the discussion that follows, the relevant operators are
\begin{align}
\mathcal{O}_9^{ii} &= \dfrac{\alpha_\mathrm{em}}{4\pi}(\bar{s}_L\gamma^\mu b_L)  (\bar{\ell}_i\gamma_\mu \ell_i)\,, \\
\mathcal{O}_{10}^{ii} &= \dfrac{\alpha_\mathrm{em}}{4\pi}(\bar{s}_L\gamma^\mu b_L)  (\bar{\ell}_i\gamma_\mu \gamma_5 \ell_i) \,,
\end{align}

\noindent as well as the primed operators, $\mathcal{O}_{9,10}^{\prime}$, which are obtained from those above by the chirality flip $P_L \leftrightarrow P_R$ in the quark current. We will not consider the electromagnetic dipole operator, $\mathcal{O}_7$, since it contributes equally to decays to electrons and muons~\cite{Altmannshofer:2008dz}. Moreover, (pseudo)scalar operators are not relevant to our discussion since they are tightly constrained by $\mathcal{B}(B_s\to\mu\mu)$ \cite{Hiller:2014yaa}, while tensor operators are forbidden at dimension-6 by the SM gauge symmetry~\cite{Buchmuller:1985jz,Alonso:2014csa}. In this section, we will omit the dependence on the renormalisation scale and take $C_a \equiv C_a (m_b)$. Effects related to operator mixing via RGEs will be discussed in Sec.~\ref{sec:loop-level}.

\begin{table*}[htbp!]
\renewcommand{\arraystretch}{2.15}
\centering
\begin{tabular}{|cc|c|c|c|c|c|}
\hline 
\;\;SMEFT\;\; & Flavour indices & Low energy WCs & Best fit & $1\sigma$ & $2\sigma$ & Pull\\ \hline \hline
\;\;$C_{\substack{lq}}^{(1,3)}$ & (2223) & $C_9^{\mu\mu}=-C_{10}^{\mu\mu}$ & $+0.30$ & $(0.18,0.40)$ & $(0.11,0.49)$ & $4.2\sigma$\\ \hline
\;\;$C_{\substack{lq}}^{(1,3)}$ & (1123) & $C_9^{ee}=-C_{10}^{ee}$ & $-0.33$ & $(-0.50,-0.20)$ & $(-0.85,-0.15)$ & $4.0\sigma$\\
\;\;$C_{{\substack{eq}}}$ & (1123) & $C_9^{ee}=C_{10}^{ee}$ & $+1.31$ & $(1.00,1.63)$ & $(0.75,1.83)$ & $4.4\sigma$\\
\;\;$C_{\substack{ed}}$ & (1123) & $(C_9^{ee})^\prime=(C_{10}^{ee})^\prime$ & $-1.36$ & $(-1.70,-1.02)\cup(1.08,1.73)$ & $(-1.90,-0.78)\cup (0.63,1.93)$ & $4.1\sigma$\\
 \hline 
\end{tabular}
\caption{ \sl \small List of viable single WCs in the SMEFT which accommodate $R_{K^{(\ast)}}$, while being consistent with $\mathcal{B}(B_s\to\mu\mu)$. The scale of new physics is considered to be $\Lambda=20~\mathrm{TeV}$. Our results are in agreement with the fits performed in Refs.~\cite{Alguero:2019ptt,Ciuchini:2019usw,Aebischer:2019mlg}.}
\label{tab:wc-tree-fit} 
\end{table*}

\subsection{Matching at the electroweak scale}
\label{ssec:tree-level}

We start by matching Eq.~\eqref{eq:leff-mb} onto the Warsaw basis \cite{Grzadkowski:2010es}, which respects the SM gauge symmetry, $SU(3)_c\times SU(2)_L\times U(1)_Y$. 
This approach is valid as long as the masses of new states are sufficiently larger than the electroweak scale, as is suggested by the status of direct searches at the LHC.  We normalise the SMEFT effective Lagrangian as
\begin{equation}
\label{eq:smeft}
\mathcal{L}_{\mathrm{SMEFT}}= \dfrac{1}{\Lambda^2}\sum_i C_i\, \mathcal{O}_i\,,
\end{equation}
where $\mathcal{O}_i$ are dimension-six operators and $C_i$ denotes their WCs introduced at the new physics scale, $\Lambda$. 
The fermionic operators in the SMEFT have definite chiralities, since they involve either left-handed or right-handed fermions.\footnote{See Appendix~\ref{app:conventions} for the conventions used in this paper.} Among the semileptonic operators, three involve left-handed quarks, 
namely\footnote{Note that we do not consider operators involving the Higgs boson and quarks only, such as $\mathcal{O}_{Hq}^{(1)}=\big{(}H^\dagger \overleftrightarrow{D}_\mu H\big{)}\left(\overline{q} \gamma^\mu q\right)$, since they induce LFU contributions to $b\to s \ell\ell$.}
\begin{align}
\label{eq:semilep-1}
	 \mathcal{O}_{\substack{eq\\prst}}  & =(\overline{e}_p \gamma_\mu e_r)  (\overline{q}_s \gamma^\mu q_t) \,,\\
\label{eq:semilep-2}
	\mathcal{O}_{\substack{lq\\prst}}^{(1)} &= (\overline{l}_p \gamma^\mu l_r)(\overline{q}_s \gamma_\mu q_t)\,,\\
\label{eq:semilep-3}
		\mathcal{O}_{\substack{lq\\prst}}^{(3)} &= (\overline{l}_p \gamma^\mu \sigma^I l_r)(\overline{q}_s \gamma_\mu \sigma^I q_t)\,,
\end{align}

\noindent where $\sigma^I$ are the Pauli matrices. These operators can be matched onto Eq.~\eqref{eq:leff-mb}  via
\begin{align}
C_9^{ii} &=  \dfrac{\pi}{\alpha_\mathrm{em} \lambda_t}\dfrac{v^2}{\Lambda^2}\left(C_{\substack{eq\\ii23}}+C_{\substack{lq\\ii23}}^{(1)}+C_{\substack{lq\\ii23}}^{(3)}\right)\,,\\
C_{10}^{ii} &=  \dfrac{\pi}{\alpha_\mathrm{em} \lambda_t}\dfrac{v^2}{\Lambda^2}\left(C_{\substack{eq\\ii23}}-C_{\substack{lq\\ii23}}^{(1)}-C_{\substack{lq\\ii23}}^{(3)}\right)\,.
\end{align}
The operators with left-handed currents, $\mathcal{O}_{lq}^{(1)}$ and $\mathcal{O}_{lq}^{(3)}$, and non-vanishing WCs for electrons and/or muons have been considered in several studies as the simplest explanation of the $R_{K^{(\ast)}}$ anomalies, cf.~e.g.~\cite{Bifani:2018zmi} for a recent review.

Another possibility is to consider operators involving right-handed quarks. While these scenarios are typically discarded as a viable explanation of the LFU hints since they cannot simultaneously explain $R_{K}^{\mathrm{exp}}<R_{K}^{\mathrm{SM}}$ and $R_{K^{\ast}}^{\mathrm{exp}}<R_{K^{\ast}}^{\mathrm{SM}}$ via new physics couplings to muons, this can be achieved in some cases if couplings to electrons are considered instead~\cite{DAmico:2017mtc}. The relevant SMEFT operators are
\begin{align}
\mathcal{O}_{\substack{ed\\prst}} &=(\overline{e}_p \gamma^\mu e_r)(\overline{d}_s \gamma_\mu d_t)\,,\\
\mathcal{O}_{\substack{ld\\prst}} &=(\overline{l}_p \gamma^\mu l_r)(\overline{d}_s \gamma_\mu d_t)\,.
\end{align}

\noindent These can be matched onto Eq.~\eqref{eq:leff-mb} via
\begin{align}
\left(C_9^{ii}\right)^\prime &= \dfrac{\pi}{\alpha_\mathrm{em} \lambda_t}\dfrac{v^2}{\Lambda^2}\left( C_{\substack{ed\\ii23}} + C_{\substack{ld\\ii23}}\right)\,,\\
\left(C_{10}^{ii}\right)^\prime &= \dfrac{\pi}{\alpha_\mathrm{em} \lambda_t}\dfrac{v^2}{\Lambda^2}\left( C_{\substack{ed\\ii23}} - C_{\substack{ld\\ii23}}\right)\,.
\end{align}

\noindent As will be discussed below, these operators require a smaller new-physics scale and/or larger couplings than purely left-handed operators to explain the present anomalies, but they nevertheless remain consistent with existing bounds.

\subsection{Tree-level explanations of the LHCb anomalies}
\label{ssec:fit-exp}

We shall now identify the effective coefficients among those of Sec.~\ref{ssec:tree-level} capable of explaining at tree-level the current deviations measured by LHCb. 
The most recent LHCb determinations of $R_{K^{(\ast)}}$~\cite{Aaij:2017vbb,Aaij:2019wad} are\footnote{Belle also performed similar LFU tests~\cite{Abdesselam:2019wac}, however we have explicitly checked that their experimental uncertainties remain too large to provide a meaningful modification of our low-energy fit.}
\begin{align}
R_K^{[1,6]} &= 0.846\,_{-0.054}^{+0.060}\,_{-0.014}^{+0.016}\,,\\[0.2em]
R_{K^\ast}^{[1.1,6]} &= 0.660 \,_{-0.070}^{+0.110}\,\pm 0.024 \,,\\[0.2em]
R_{K^\ast}^{[0.045,1.1]} &= 0.685\,_{-0.069}^{+0.113}\,\pm 0.047\,.
\end{align}

\noindent Moreover, a weighted average of the latest LHCb, CMS and ATLAS measurements~\cite{Aaij:2017vad,Chatrchyan:2013bka,Aaboud:2018mst} gives
\begin{equation}
\mathcal{B}(B_s\to \mu\mu)^{\mathrm{exp}}=(2.93\pm 0.42)\times 10^{-9}\,.
\end{equation}
This branching ratio is the cleanest observable related to the transition $b\to s\mu\mu$, as far as hadronic uncertainties are concerned, and it is slightly below, though still in reasonable agreement with, the SM prediction, $\mathcal{B}(B_s\to \mu\mu)^{\mathrm{SM}}=(3.65\pm 0.23)\times 10^{-9}$~\cite{Bobeth:2013uxa}, given the large uncertainties.
In our phenomenological analysis, we prefer to focus on the observables listed above, since the theoretical predictions for other $b\to s\ell\ell$ quantities can be affected by hadronic uncertainties which are not yet under full theoretical control~\cite{Becirevic:2011bp}. In scenarios with new physics coupled to muons, it is important to stress that our results are in reasonable agreement with the ones from the global analyses~\cite{Alguero:2019ptt,Ciuchini:2019usw,Aebischer:2019mlg}.

In Table~\ref{tab:wc-tree-fit}, we list the single WCs which can provide a significantly improved description of current data via a tree-level contribution, along with their best-fit regions. Flavour indices are chosen to produce tree-level contributions, assuming that the Yukawa matrix is diagonal in the down-quark sector. 
The scale of new physics is fixed for illustration to be $\Lambda= 20$~TeV. 
The successful scenarios are chosen by requiring that the pull for a single degree of freedom, $\sqrt{\chi^2_\mathrm{SM}-\chi^2_{\text{best fit}}}$, gives at least a $3\sigma$ improvement on the SM.

We considered the range $|C_i/\Lambda^2| \lesssim 1/(10{~\rm TeV})^2$ for the WCs in Table~\ref{tab:wc-tree-fit}, so that the new physics contributions to $R_{K^{(\ast)}}$ are sub-dominant with respect to the SM ones. This requirement allows us to discard far-fetched solutions that involve a large cancellation between the SM and new physics contributions.
From Table~\ref{tab:wc-tree-fit}, we see that  the present discrepancies can be accommodated with left-handed operators satisfying
\begin{equation}
\label{eq:lh-solution}
C_{\substack{lq\\2223}}^{(1,3)} - C_{\substack{lq\\1123}}^{(1,3)}  \simeq 0.3 \times \left(\frac{\Lambda}{20~\mathrm{TeV}}\right)^2 \,,
\end{equation}
where the new physics contribution can arise via the couplings to electrons or muons.\footnote{As an example of a solution involving a large cancellation, we mention that an equally good fit is realised by replacing $\simeq 0.3$ with $\simeq {-5.7}$ in Eq.~\eqref{eq:lh-solution}, and by setting the muonic coupling to be zero.} 
More importantly, as already anticipated in the previous section, we find viable solutions with couplings to right-handed electrons. Note, in particular, that these scenarios require a new physics WC about four times larger than the ones in Eq.~\eqref{eq:lh-solution}.

\begin{figure*}[t!]
\centering
\includegraphics[width=0.495\textwidth]{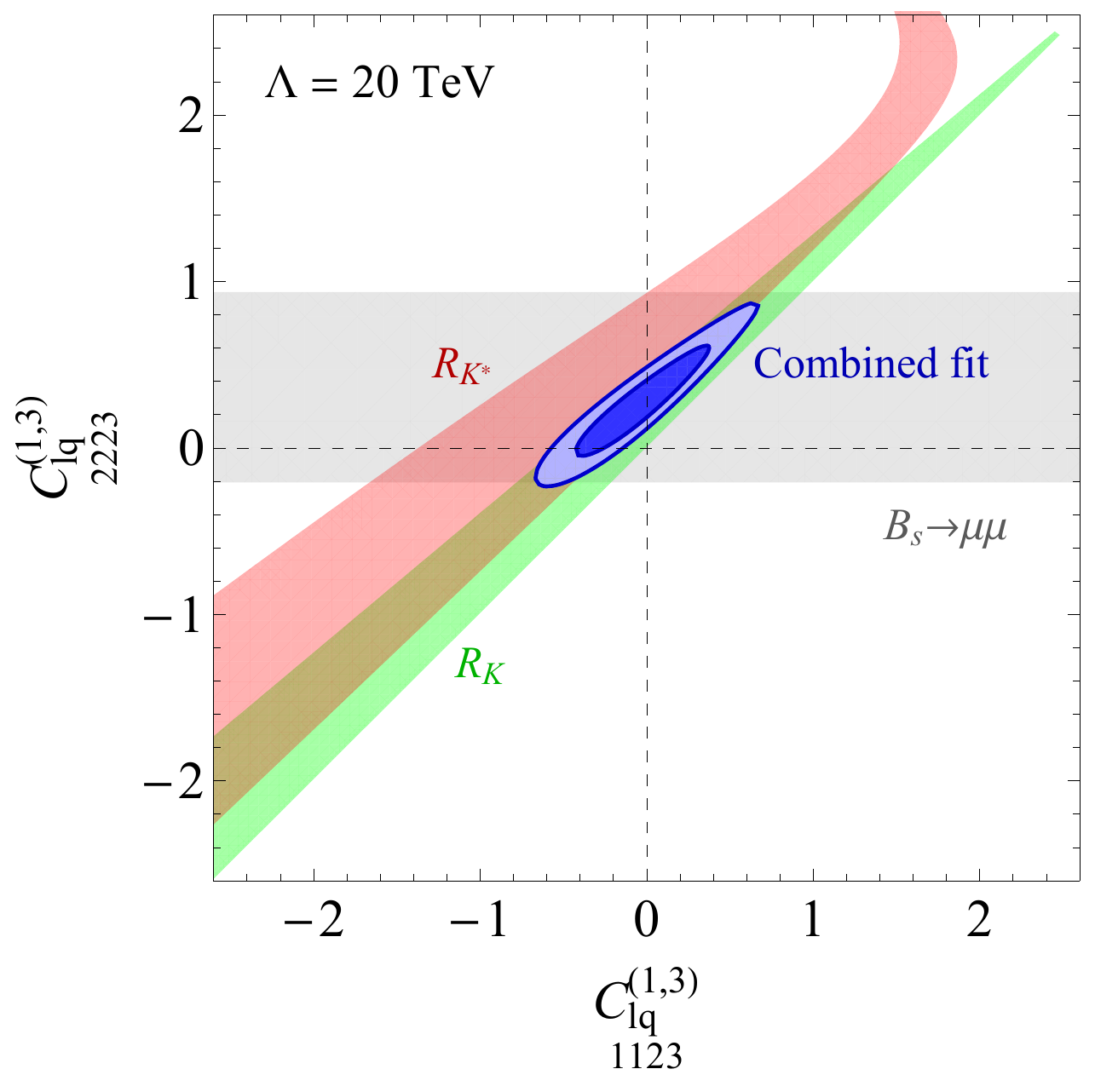}
\includegraphics[width=0.495\textwidth]{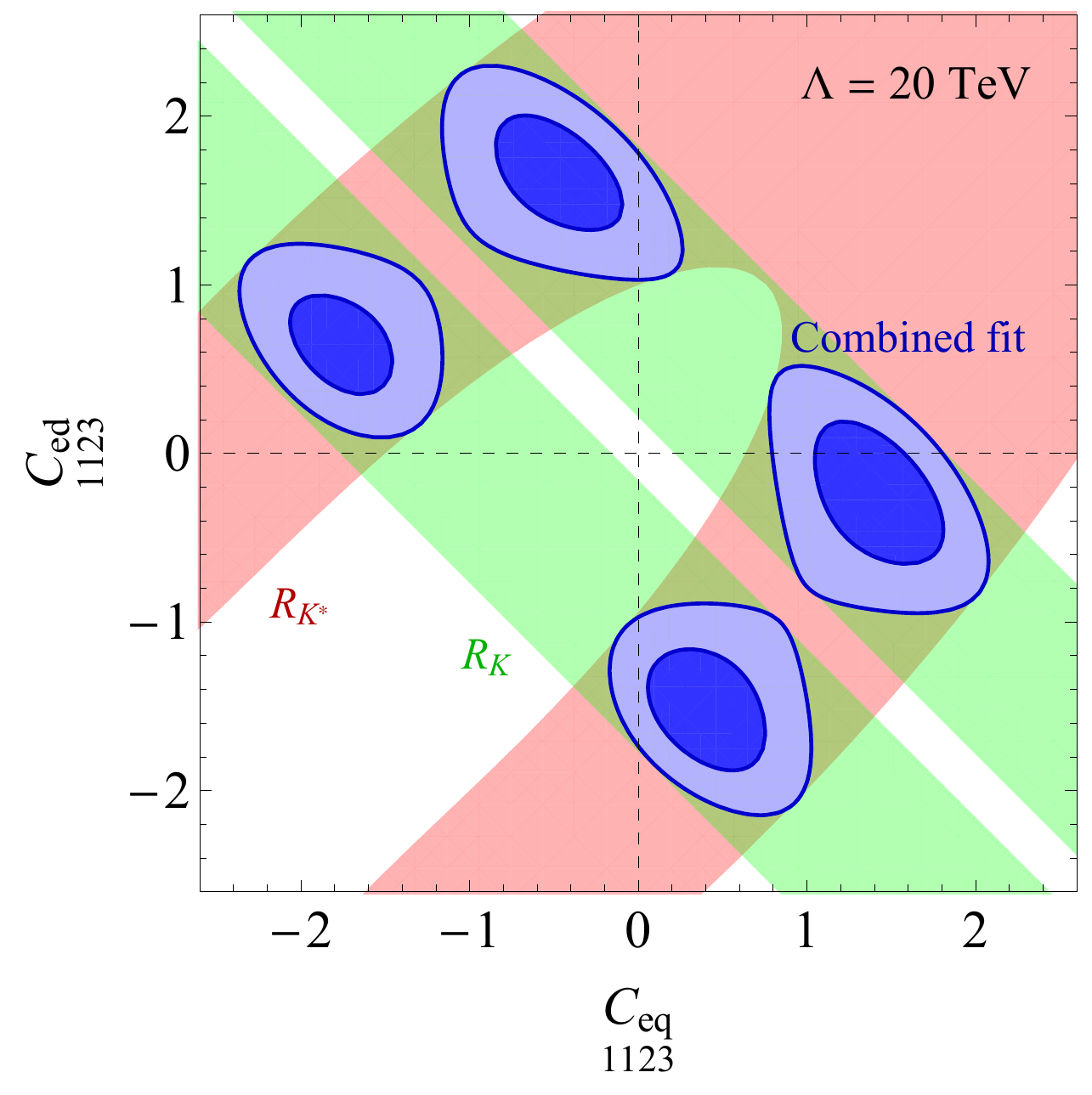}
\caption{\sl Constraints on the left-handed WCs $C_{lq}^{(1,3)}$ coupled to electrons and muons (left panel), and the WCs containing right-handed lepton currents, $C_{eq}$ and $C_{ed}$, coupled to electrons (right panel). 
The allowed regions are derived from the experimental measurements to $2\sigma$ accuracy of $R_K$ (green), $R_{K^\ast}$ (red) and $B_s\to\mu\mu$ (gray) by assuming $\Lambda=20$~TeV. The combined fit of this data is shown by the dark (light) blue regions at $1\sigma$~$(2\sigma)$.} 
\label{fig:fits-tree}
\end{figure*}

To further illustrate our results, two scenarios of pairs of WCs are shown in Fig.~\ref{fig:fits-tree}: (i) operators with purely left-handed currents, $\mathcal{O}_{lq}^{(1,3)}$, coupled to electrons and muons (left panel), and (ii) right-handed lepton operators, $\mathcal{O}_{eq}$ and $\mathcal{O}_{ed}$, with couplings only to electrons (right panel). 
The only solution we find in the first scenario is the one described by Eq.~\eqref{eq:lh-solution}. 
The case of operators with right-handed lepton currents has several solutions since they contribute differently to $R_K$ and $R_{K^{\ast}}$, as shown in the right panel of Fig.~\ref{fig:fits-tree}. Some of them are achieved with a single WC, as described in Table~\ref{tab:wc-tree-fit}. 

The above discussion considers only WCs generated at tree-level at the scale $\Lambda$. 
However, non-negligible contributions can also arise at loop level. Loop effects may be the main source of lepton flavour universality violation, or they can appear on top of tree-level contributions, when a more general flavour structure is considered, as we shall discuss now.

\section{Effective theory at one loop}
\label{sec:loop-level}

In this section we extend our discussion to LFU violation effects generated through renormalisation group evolution from the ultraviolet scale, $\Lambda$, down to the scale of $B$-physics experiments, 
$\mu \simeq m_b$. SM interactions induce non-trivial operator mixing from $\Lambda$ down to the electroweak scale, which we identify for definiteness as the top-quark mass, $\mu_{\mathrm{EW}} \simeq m_t$, thus neglecting the small difference between $m_t$ and $m_W$. 
The RGE contributions below the electroweak scale are negligible, since the QCD corrections vanish for semi-leptonic operators with a (axial-)vector quark current, which are protected by the Ward identity.

We will now classify the operators that do not contribute to $R_{K^{(\ast)}}$ at tree-level, but rather via one-loop diagrams, and quantify these contributions. 
These are scenarios which generate one of the operators identified in Table \ref{tab:wc-tree-fit} at loop level. 
To identify the viable scenarios, we consider a leading-logarithmic approximation in our analytical expressions. 
Only the dominant RGE effects will be kept, namely those proportional to the top-quark or charm-quark Yukawas, provided that the latter are enhanced by a CKM factor (e.g.~$\propto V_{cs}/V_{ts}$). 
Loops involving other Yukawa couplings can safely be ignored. 
Contributions induced by the bottom-quark Yukawa (i.e.~the largest Yukawa we neglect) cannot be CKM enhanced and are therefore sub-dominant. Gauge loops can also be neglected, as they do not change the operator flavour and chirality structure, as required to obtain a one-loop contribution to $R_{K^{(\ast)}}$.
The validity of these approximations has been corroborated by using a numerical code which accounts for one-loop RGE effects~\cite{Straub:2018kue}. Finite (non-logarithmically enhanced) one-loop effects cannot be extracted from our RGE analysis, but we will point out some cases where they may be relevant.\footnote{See Ref.~\cite{Aebischer:2015fzz,Hurth:2019ula,Dekens:2019ept} for one-loop matching results in the EFT of $b \to s$ transitions.} 
Two-loop contributions can be safely neglected, as they are sizeable only for $\Lambda$ below the electroweak scale, which is forbidden by a number of experimental constraints.

\subsection{SMEFT operators mixing into $R_{K^{(\ast)}}$}
\label{ssec:SMEFT}

Loop contributions to $R_{K^{(\ast)}}$ could arise from two different sources:
\begin{itemize}
\item[(a)] Operators with a different Lorentz and/or gauge structure to the SMEFT operators which contribute at tree-level, listed in Sec.~\ref{ssec:tree-level}.
\item[(b)] Operators with the same Lorentz and gauge structure as the tree-level ones, but with a choice of flavour indices that forbids tree-level contributions.\footnote{Recall that we define SMEFT operators in a basis where $Y_d$ is diagonal at the scale $\Lambda$.}
\end{itemize}

For scenario (a), keeping our assumptions on the Yukawa dominance of the RGE contributions, we find that the new operators that mix via RGEs into those listed in Sec.~\ref{ssec:tree-level} are
\begin{align}
\label{eq:OHe}
\mathcal{O}_{\substack{He \\ii}} &= \left(H^\dagger \overleftrightarrow{D}_\mu H\right)\left(\overline{e}_i \gamma^\mu e_i\right)\,,\\
\label{eq:OHl1}
\mathcal{O}_{\substack{Hl \\ii}}^{(1)} &=\left(H^\dagger \overleftrightarrow{D}_\mu H\right)\left(\overline{l}_i \gamma^\mu l_i\right)\,,\\
\label{eq:OHl3}
\mathcal{O}_{\substack{Hl \\ii}}^{(3)} &= \left(H^\dagger \overleftrightarrow{D}_\mu \sigma^I H\right)\left(\overline{l}_i \gamma^\mu  \sigma^I l_i\right)\,,
\end{align}
with flavour indices $i\in \lbrace 1,2\rbrace$, and the semileptonic operators
\begin{align}
\label{eq:semilep-bis-1}
\mathcal{O}_{\substack{eu\\ iist}}&= (\overline{e}_i \gamma^\mu e_i)(\overline{u}_s \gamma_\mu u_t)\,, \\
\label{eq:semilep-bis-2}
\mathcal{O}_{\substack{lu\\ iist}}&= (\overline{l}_i \gamma^\mu l_i)(\overline{u}_s \gamma_\mu u_t)\,,
\end{align}
where the dominant effects come from flavour indices  $(s,t)=(2,3)$ or $(3,3)$.

For scenario (b), one should consider the operators of Sec.~\ref{ssec:tree-level}, but with different quark flavour indices.
More specifically, the relevant possibilities are
\begin{equation*}
\mathcal{O}_{\substack{eq\\ iist}}\,,\quad  \mathcal{O}_{\substack{l q \\ iist}}^{(1,3)} \,,\quad {\rm for}~(s,t)=(2,2)~{\rm or}~ (3,3)\,.
\end{equation*}
The choice of flavour indices is meant to prevent a tree-level contribution to $R_{K^{(\ast)}}$, which requires $(s,t)=(2,3)$, and to allow for the dominant one-loop effects, namely those driven by the top-quark Yukawa. Note that the operators $\mathcal{O}_{ed}$ and $\mathcal{O}_{l d}$ cannot induce one-loop quark-flavour change in the basis where $Y_d$ is diagonal at $\Lambda$.

These potential one-loop explanations of the anomalies require a cutoff, $\Lambda$, close to the TeV scale, therefore one should carefully inspect experimental constraints from precision electroweak measurements, low energy flavour observables, and direct searches at colliders. 
Note that these constraints are much milder for tree-level contributions to $R_{K^{(\ast)}}$, as one can take $\Lambda$ above $\sim 10$ TeV.

\subsection{Experimental constraints}
\label{ssec:constraints}

There are several experimental constraints on the scenarios we consider, which we now discuss in detail.\\

\paragraph*{$Z$-pole observables.} The operators listed above induce new contributions to the leptonic $W$ and $Z$-boson couplings, which are very well constrained by LEP data~\cite{ALEPH:2005ab}. The $Z$-boson couplings can be parametrised in terms of the effective Lagrangian

\begin{equation}
\mathcal{L}_{\mathrm{eff}}^{Z}= - \dfrac{g}{\cos\theta_W} \sum _{f,i} \bar{f}_i\,\gamma_\mu \Big{[}g_{f_L}^i P_L + g_{f_R}^i P_R \Big{]}f_i \, Z^\mu\,,
\end{equation}
\noindent where $\theta_W$ is the weak mixing angle and
\begin{equation}
g_{f_{L(R)}}^i= g_{f_{L(R)}}^{\mathrm{SM}}+\delta g_{f_{L(R)}}^{\mathrm{SM}}\,,
\end{equation}
\noindent with $g_{f_L}^\mathrm{SM}=T_3^f-Q_f \, \sin^2 \theta_W$ and $g_{f_R}^\mathrm{SM}=-Q_f \, \sin^2 \theta_W$. New physics contributions are described by $\delta g^Z_{\ell_{Li}}$, which can be matched at $\mu_\mathrm{EW}$ onto the Warsaw basis via the relations
\begin{align}
\label{eq:Z-pole-1}
\delta g^Z_{\nu_{Li}} &= - \dfrac{v^2}{2\Lambda^2}\left(C_{\substack{Hl\\ ii}}^{(1)}-C_{\substack{Hl\\ ii}}^{(3)}\right)\,, \\
\label{eq:Z-pole-2}
\delta g^Z_{\ell_{Li}} &= - \dfrac{v^2}{2\Lambda^2}\left(C_{\substack{Hl\\ ii}}^{(1)}+C_{\substack{Hl\\ ii}}^{(3)}\right)\,, \\
\label{eq:Z-pole-3}
\delta g^Z_{e_{Ri}} &= - \dfrac{v^2}{2\Lambda^2}C_{\substack{He\\ ii}}\,,
\end{align}
where the WCs on the right-hand sides should be evaluated at $\mu=\mu_\mathrm{EW}$. Note that semileptonic operators, such as those listed in Eqs.~\eqref{eq:semilep-bis-1} and \eqref{eq:semilep-bis-2}, may contribute to $C_{Hl}^{(1)}$, $C_{Hl}^{(3)}$ and $C_{He}$ at the one-loop level. 
In our analysis, we consider the fit to LEP data performed in Ref.~\cite{Efrati:2015eaa}, which accounts for the correlation among $Z$ and $W$ couplings to leptons arising from $SU(2)_L\times U(1)_Y$ gauge invariance. 
We also performed our own, independent analysis and found good agreement with the results of Ref.~\cite{Efrati:2015eaa}.

For illustration, we quote the constraints on $C_{Hl}^{(1)}\pm C_{Hl}^{(3)}$ for muons at $2\sigma$ accuracy, derived from the ensemble of $Z$-pole observables and evaluated at $\mu_\mathrm{EW}$. We have
\begin{align}
\label{eq:Z-pole-example}
C_{\substack{Hl\\ 22}}^{(1)} + C_{\substack{Hl\\ 22}}^{(3)} &\in \left(-2.4,1.0\right)  \times 10^{-2} \left(\dfrac{\Lambda}{1~\mathrm{TeV}}\right)^2\,,\\[0.4em]
C_{\substack{Hl\\ 22}}^{(1)} - C_{\substack{Hl\\ 22}}^{(3)} &\in \left(0.1, 1.4\right)  \times 10^{-1} \left(\dfrac{\Lambda}{1~\mathrm{TeV}}\right)^2\,,
\label{eq:Z-pole-example-2}
\end{align}
with a strong correlation in the plane $C_{Hl}^{(1)}$ vs.~$C_{Hl}^{(3)}$. The latter combination, with the minus sign, is subject to a weaker bound since the $Z$-couplings to neutrinos are less constrained than those to charged leptons, cf.~Eqs.~\eqref{eq:Z-pole-1} and \eqref{eq:Z-pole-2}.

\

\paragraph*{LFU in kaon decays.}

The operators $\mathcal{O}_{lq}^{(1)}$ and $\mathcal{O}_{lq}^{(3)}$, defined in Eq.~\eqref{eq:semilep-2} and \eqref{eq:semilep-3}, are constrained by LFU tests in tree-level semileptonic decays. The most stringent limit arises from the ratio defined as  
\begin{equation}
r_{K}^{e/\mu} = \dfrac{\mathcal{B}(K \to e \bar{\nu})}{\mathcal{B}(K \to \mu \bar{\nu})}\,,
\end{equation}
for which the experimental measurement gives $r_K^{e/\mu\,(\mathrm{exp})}=(2.488\pm 0.010)\times 10^{-5}$ \cite{Tanabashi:2018oca}, in good agreement with the SM prediction, $r_K^{e/\mu\,(\mathrm{SM})}=(2.477\pm 0.001)\times 10^{-5}$ \cite{Cirigliano:2007xi}. Among the WCs relevant for $R_{K^{(\ast)}}$, those with flavour indices $ii22$ receive the strongest constraint from this observable as they depend on the same CKM elements as the SM amplitude. More explicitly, we obtain
\begin{equation}
\dfrac{r_K^{e/\mu\,(\mathrm{exp})}}{r_K^{e/\mu\,(\mathrm{SM})}} \approx 1-\dfrac{2\,v^2}{\Lambda^2}\left(C_{\substack{lq\\ 1122}}^{(3)}-C_{\substack{lq\\ 2222}}^{(3)}\right)\,,
\end{equation}
where the running effects have been neglected for simplicity\footnote{The electroweak running between $\mu=\Lambda$ and $\mu_{\mathrm{EW}}$ can amount to $\approx 20\%$ corrections, while the one below $\mu_{\mathrm{EW}}$ is entirely negligible~\cite{Gonzalez-Alonso:2017iyc}. These effects are included in our numerical analysis.}. From this expression, we obtain the constraint
\begin{equation}
\label{eq:kaon-lfuv}
 C_{\substack{lq\\ 1122}}^{(3)}-C_{\substack{lq\\ 2222}}^{(3)} \in (-0.10,0.03)\times\left(\dfrac{\Lambda}{1~\mathrm{TeV}}\right)^2\,.
\end{equation}
Note, also, that $\mathcal{O}_{lq}^{(1)}$ contributes to a shift in $r_K^{e/\mu}$ only at one loop, hence the bounds on its WCs are correspondingly weaker.

\paragraph*{LFU in $B$-meson decays.}

Similarly, important constraints arise from LFU tests in $B$-meson decays, namely
 
\begin{equation}
R_{D}^{\mu/e} = \dfrac{\mathcal{B}(B \to D \mu \bar{\nu})}{\mathcal{B}(B \to D e \bar{\nu})}\,,
\end{equation}

\noindent which was experimentally determined as $R_D^{\mu/e}=0.995(22)(39)$~\cite{Glattauer:2015teq}, in agreement with the SM prediction $R_D^{\mu/e}= 0.9957(4)$, obtained by using the lattice QCD form factors from Refs.~\cite{Lattice:2015rga,Na:2015kha}. As a consequence, we find
\begin{equation}
 C_{\substack{lq\\ 2233}}^{(3)}-C_{\substack{lq\\ 1133}}^{(3)} \in (-0.70,0.80) \times\left(\dfrac{\Lambda}{1~\mathrm{TeV}}\right)^2\,.
\end{equation}
These bounds are weaker than those derived from kaon decays, cf.~Eq.~\eqref{eq:kaon-lfuv}, but they have the advantage of being sensitive to third-generation quark couplings.

\

\paragraph*{Collider bounds on contact interactions.} Relevant experimental constraints on effective operators with electrons can be extracted from LEP limits on $\sigma(e^+ e^- \to q_i\bar{q}_j)$ obtained at center-of-mass energies as large as $\sqrt{s}=209$~GeV~\cite{Abbiendi:2001wk,Schael:2006wu}. The most stringent limits on flavour-violating operators comes 
from the combined LEP data~\cite{Aleph:2001dzz}, from which we find, for the relevant channel $\sigma(e^+ e^-\to c\bar{t})$,
\begin{equation}
\left|C_{\substack{\alpha\\ 1123}}\right| \lesssim 1.5 \times \left(\dfrac{\Lambda}{1~\mathrm{TeV}}\right)^2\,,
\label{lepctbound}
\end{equation}
where $\alpha \in \lbrace lq^{(1,3)}, lu, eq, eu \rbrace$, see also Ref.~\cite{BarShalom:1999iy}. For flavour-conserving operators, we obtain the most stringent limits for  $\sigma(e^+ e^-\to b\bar{b} \ / \ c\bar{c} \ / \ u\bar{u}+d\bar{d}+s\bar{s} )$ 
from ALEPH data~\cite{Schael:2006wu}, which allows us to constrain operators with $\Lambda \approx 1$~TeV and $\mathcal{O}(1)$ couplings.

A bound can also be placed on operators contributing to the decays $t \to c \ell \ell$, where $\ell = e,\mu$. 
ATLAS sets the upper limit $\mathcal{B}(t \to c Z) < 2.4 \times 10^{-4}$ at $95\%$ C.L. \cite{Aaboud:2018nyl}, by selecting $Z$ decays into electrons and muons with dilepton invariant mass in the window $m_{\ell \ell} \in [m_Z - 15\text{ GeV}, m_Z + 15\text{ GeV}]$.
Adding to the SM an operator $\frac{C_{XY}}{\Lambda^2} (\overline{\ell} \gamma_\nu P_X \ell) (\overline{c} \gamma^\nu P_Y t)$, with $X,Y$ being either $L$ or $R$, we find 
\begin{equation}
\dfrac{\mathrm{d}\mathcal{B}(t\to c \ell\ell)}{\mathrm{d}m_{\ell\ell}^2}= \dfrac{ m_t^3\,|C_{XY}|^2}{768 \pi^3\Lambda^4 \Gamma_t}\left(1- 3 x_t^2+ 2 x_t^3\right)\,,
\end{equation}
where $x_t= m_{\ell\ell}^2/m_t^2$ and $\Gamma_t$ is the top-quark width. Integration over $m_{\ell\ell}$ then gives
\begin{equation}
\mathcal{B}(t \to c \ell \ell)_{m_{\ell\ell}\in[m_Z\pm15{\rm GeV}]} \simeq 0.3\, \frac{ m_t^5\, |C_{XY}|^2}{1536 \pi^3 \Lambda^4 \Gamma_t}~,
\end{equation}
where the factor $\simeq 0.3$ comes from the restriction on the dilepton invariant mass.
Since the ATLAS bound is obtained by combining electron and muon events, we obtain
\begin{equation}
\left|C_{\substack{\alpha\\ ii23}} \right| \lesssim 5.1 \times \left( \frac{\Lambda}{1 \text{ TeV}} \right)^2~,
\end{equation}
for $i=1,2$, where $\alpha$ takes the values given just after Eq.~\eqref{lepctbound}. 
For operators with electrons, this is weaker than the LEP bound discussed above, but for several operators with muons it constitutes the strongest constraint on the WC, see Table \ref{tab:wc-pred-loop}. 
Note that our naive recast of the ATLAS bound might change if this experimental analysis were optimised for the $tc\ell\ell$ contact interactions, since these operators contribute to the same final state of the ATLAS analysis via $pp (g g) \to tc\mu\mu$. However, a complete LHC analysis lies beyond the scope of this paper.

Finally, we comment on similar bounds on contact interactions which can be derived from high-$p_T$ dilepton tails at the LHC~\cite{Faroughy:2016osc,Greljo:2017vvb}. 
While stringent limits can be derived from this data, one should be cautious about the EFT's validity. 
Given the current experimental precision, one can probe four-fermion operators with scales $\Lambda \simeq \mathcal{O}(1~\mathrm{TeV})$. However, since LHC analyses observe events up to invariant dilepton mass $m_{ll} \sim \mathcal{O}(3)$~TeV \cite{Aaboud:2017buh}, the EFT description breaks down. Thus, unlike for our treatment of LEP data, one should specify the propagating degree of freedom, i.e.~the mediator and its couplings, in order to correctly assess the limits in this case. 
We will address this issue in Sec.~\ref{sec:simplified}.

\begin{table*}[htbp!]
\renewcommand{\arraystretch}{2.2}
\centering
\begin{tabular}{|c|c|c|c|}
\hline
$\begin{array}{c}{\rm Wilson}\\[-10pt]{\rm Coefficients}\end{array}$  & $\begin{array}{c}{\rm Flavour}\\[-10pt]{\rm Indices}\end{array}$
&  $2\sigma$ range & $R_{K}$ \\ \hline \hline
 & $(2222)$ & $(-0.03,0.10)$ &  $\approx 1$   \\ 
 \multirow{-2}{*}{\;\;$C_{\substack{lq}}^{(3)}$} & $(2233)$ & $(-0.60,0.24)$ &  $(0.95,1.13)$   \\ \hline
\rowcolor{Gray} & $(2222)$ & $(-5.4,0.90)$ & $(0.48,1.1)$  \\ 
\rowcolor{Gray} \multirow{-2}{*}{\;\;$C_{\substack{lq}}^{(1)}$} & $(2233)$ & $(-0.31,0.72)$ & $(0.85,1.07)$  \\ \hline
  & $(2222)$ & $(-0.03,0.10)$ & $(0.99,1.03)$ \\ 
\rowcolor{Gray} \multirow{-2}{*}{\;\;$C_{\substack{lq}}^{(1)}=C_{\substack{lq}}^{(3)}$} & $(2233)$ & $
(-0.56,0.42)$ & $(0.83,1.25)$  \\ \hline
  & $(2222)$ & $(-1.92,10)$ &  $\approx 1$  \\ 
  \multirow{-2}{*}{\;\;$C_{\substack{eq}}$} & $(2233)$ & $(-0.90,0.24)$ & $\approx 1$  \\ \hline
\rowcolor{Gray} & $(2223)$ & $(-5.1,5.1)$ & $(0.94,1.06)$  \\ 
\rowcolor{Gray} \multirow{-2}{*}{\;\;$C_{\substack{lu}}$} & $(2233)$ & $(-0.76,0.36)$ & $(0.92,1.04)$ \\ \hline
 & $(2223)$ & $(-5.1,2.4)$ &  $ (1,1.02)$  \\ 
\multirow{-2}{*}{\;\;
$C_{\substack{eu}}$} & $(2233)$ & $(-0.28,0.96)$ &  $\approx 1$  \\ \hline
\;$C_{He}$,\, $C_{Hl}^{(1)}$\, or\, $C_{Hl}^{(3)}$ & $(22)$ & $(-0.04,0.05)$ & $\approx 1$  \\ 
\;\;$C_{Hl}^{(1)}=-C_{Hl}^{(3)}$ & $(22)$ & $(0.0,0.13)$ &  $\approx 1$ \\ \hline 
\end{tabular}
\hspace*{2.3em}
\begin{tabular}{|c|c|c|c|}
\hline
$\begin{array}{c}{\rm Wilson}\\[-10pt]{\rm Coefficients}\end{array}$  & $\begin{array}{c}{\rm Flavour}\\[-10pt]{\rm Indices}\end{array}$
 &  $2\sigma$ range& $R_{K}$ \\ \hline \hline
 & $(1122)$ & $(-0.10,0.02)$ &  $\approx 1$  \\ 
 \multirow{-2}{*}{\;\;$C_{\substack{lq}}^{(3)}$} & $(1133)$ & $(-0.05,0.48)$ &  $(0.98,1.11)$  \\ \hline
 & $(1122)$ & $(-0.19,0.14)$ & $\approx 1$ \\ 
\rowcolor{Gray} \multirow{-2}{*}{\;\;$C_{\substack{lq}}^{(1)}$} & $(1133)$ & $(-0.41,0.02)$ & $(0.91,1.01)$  \\ \hline
  & $(1122)$ & $(-0.10,0.03)$ & $\approx 1$  \\ 
\rowcolor{Gray} \multirow{-2}{*}{\;\;$C_{\substack{lq}}^{(1)}=C_{\substack{lq}}^{(3)}$} & $(1133)$ & $(-0.41,0.18)$ & $(0.84,1.09)$ \\ \hline
  & $(1122)$ & $(-0.35,0.83)$ & $\approx 1$   \\ 
  \multirow{-2}{*}{\;\;$C_{\substack{eq}}$} & $(1133)$ & $(-0.21,0.28)$ & $\approx 1$ \\ \hline
 & $(1123)$ & $(-1.5,1.5)$ & $(0.97,1.02)$  \\ 
\multirow{-2}{*}{\;\;$C_{\substack{lu}}$} & $(1133)$ & $(-0.02,0.43)$ & $(0.95,1.01)$  \\ \hline
 & $(1123)$ & $(-1.5,1.5)$ &  $\approx 1$  \\ 
\multirow{-2}{*}{\;\;
$C_{\substack{eu}}$} & $(1133)$ & $(-0.29,0.21)$ &  $\approx 1$  \\ \hline
\;$C_{He}$,\, $C_{Hl}^{(1)}$\, or\, $C_{Hl}^{(3)}$ & $(11)$ & $(-0.02,0.03)$ & $\approx 1$ \\ 
\;\;$C_{Hl}^{(1)}=-C_{Hl}^{(3)}$ & $(11)$ & $(-0.03,0.02)$ &  $\approx 1$  \\ \hline 
\end{tabular}
\caption{ \sl \small Allowed range of values of the WCs and of $R_K \equiv R_K^{[1,6]}$ for each of the operators listed in Sec.~\ref{ssec:SMEFT}, after imposing all the constraints listed in Sec.~\ref{ssec:constraints} except for the LHC contact interaction bounds. 
The range of $R_{K^*}$ in the central $q^2$ bin is virtually identical to the $R_K$ range. 
We fix $\Lambda=1$~TeV and enforce $|C| \leq 10$, which corresponds to $|\delta C_{9,10}| \lesssim |C_{9,10}^{\mathrm{SM}}|$, where relevant. 
The selection of quark flavour indices is explained in the text. 
The WCs that can accommodate a deviation in $R_{K^{(*)}}$ of more than $5\%$ are shaded in grey. 
These give an individual pull against the SM between $\sim 2.5\sigma$ and $\sim 4\sigma$, depending on the operator.}
\label{tab:wc-pred-loop} 
\end{table*}

\subsection{Numerical results}
\label{ssec:numerical}

Now we turn to an estimate of the loop contributions to $R_{K^{(\ast)}}$ from the operators listed above. We used the numerical code \emph{flavio}~\cite{Straub:2018kue}, combined with the package \emph{Wilson}~\cite{Aebischer:2018bkb} for the matching and running of effective coefficients above the electroweak scale.~\footnote{We have also performed cross-checks of our analytical computation with the DsixTools package~\cite{Celis:2017hod}.}
We have verified these numerical results by explicitly computing the RGE effects from the anomalous-dimension matrices given in Ref.~\cite{Jenkins:2013zja,Jenkins:2013wua,Alonso:2013hga} at leading-log approximation, as we discuss below. 
We have further confirmed that one-loop matching effects computed in \cite{Aebischer:2015fzz,Hurth:2019ula} do not qualitatively change our results.

Our results are summarised in Table~\ref{tab:wc-pred-loop}, where we give the maximal deviation in $R_K\approx R_{K^\ast}$, in the $q^2 \in [1,6]~\mathrm{GeV}^2$ bin, for each operator listed in Sec.~\ref{ssec:SMEFT} after enforcing the constraints discussed in \ref{ssec:constraints}. 
Specifically, we impose that the WC gives a pull away from the SM of no more than $2\sigma$ with respect to $Z$-pole and LFU meson decay bounds, and simultaneously respects the contact interaction limits set by LEP at $95\%$ C.L.
It should be stressed that we work in the basis where $Y_d$ is diagonal at $\mu=\Lambda$ and then we rediagonalise at $\mu=\mu_{\mathrm{EW}}$, since we are interested in down-quark FCNC effects. Accounting for the misalignment of the Yukawa matrix induced at one loop has a sizeable impact on the predictions for operators containing quark doublets, as we will show in Sec.~\ref{ssec:Olq}.

From Table~\ref{tab:wc-pred-loop}, we observe that there are a few scenarios which can produce deviations in $R_{K^{(\ast)}}$ between $\mathcal{O}(5\%)$ and $\mathcal{O}(50\%)$. One of these operators is $\mathcal{O}_{lu}$ with couplings to muons, as already pointed out in Refs.~\cite{Celis:2017doq,Camargo-Molina:2018cwu}. In our analysis, we observe for the first time that $\mathcal{O}_{lq}^{(1)}$ and $\mathcal{O}_{lq}^{(1)}+\mathcal{O}_{lq}^{(3)}$ can accommodate even larger deviations for certain flavour indices. 
Note that there are more successful cases for operators with muons than with electrons, since the latter face additional constraints from LEP with respect to the former. 
We also note that operators containing a Higgs current can only induce very small effects, since they are constrained at tree-level by $Z$-pole observables. 

\section{Viable one-loop scenarios}
\label{sec:viable}

We shall now discuss in detail the two main viable scenarios. 
This will allow us to discuss the general features of the possibilities listed in Sec.~\ref{ssec:SMEFT}, as well as to retrospectively justify the choice of flavour indices in our numerical analysis.

\subsection{\underline{$\mathcal{O}_{lu} = (\bar{l} \gamma^\mu l)(\bar{u} \gamma_\mu u)$}}
\label{ssec:Olu}

\begin{figure}[h!]
\centering
\includegraphics[width=0.5\textwidth]{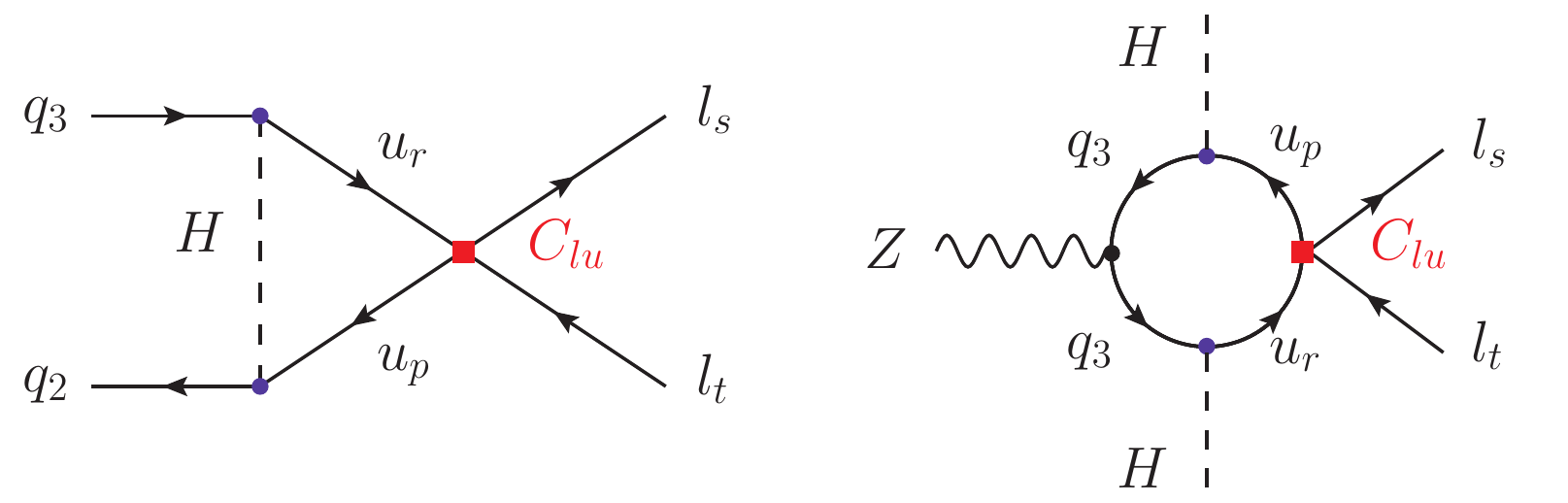}
\caption{\sl Diagrams contributing to $\mathcal{O}_{lq}^{(1)}$ (left) and $\mathcal{O}_{Hl}^{(1)}$ (right) via the running of $O_{lu}$. Only the contributions proportional to Yukawa couplings are shown, and flavour indices are denoted by $p,r,s,t$. Below the EWSB scale, these diagrams induce contributions to the $b\to s\ell\ell$ transition and to $Z$-boson couplings to leptons, respectively.}
\label{fig:diagrams-Olu}
\end{figure}

The first example we consider is the operator $\mathcal{O}_{lu}$, defined in Eq.~\eqref{eq:semilep-bis-2}. Even though this operator does not contribute to FCNCs in the down-quark sector at tree-level, it induces contributions at one loop, as depicted in Fig.~\ref{fig:diagrams-Olu}. By considering the RGE running from $\mu=\Lambda$ to $\mu_{\mathrm{EW}}$, and keeping the dominant terms, we find that the Lagrangian at $\mu = \mu_{EW}$ describing semileptonic processes contains,
\begin{equation}
{\cal L}_{\text{SMEFT}}\supset \dfrac{\log \left(\Lambda/m_t\right)}{16 \pi^2 \Lambda^2}\, C_{\substack{lu \\ prvw}}\, [Y_u^\dagger]_{sv}\,[Y_u]_{wt}\, \mathcal{O}_{\substack{lq\\prst}}^{(1)}\,,
\end{equation}

\noindent where $Y_u$ denotes the up-type quark Yukawa, defined in Appendix~\ref{app:conventions}, and $\mathcal{O}_{lq}^{(1)}$ is defined in Eq.~\eqref{eq:semilep-2}. By keeping the dominant terms in the above expression, we find that the WCs at $\mu=m_b$ read
\begin{align}
\label{eq:C9mC10-Olu}
\begin{split}
C_9^{pr} &= - C_{10}^{pr}\\[0.3em]
 &\simeq \dfrac{v^2\,\log \left(\Lambda/m_t\right)}{16 \pi^2 \Lambda^2}\,\dfrac{\pi \,y_t^2}{\alpha_{\mathrm{em}}} \,\Bigg{[}C_{\substack{lu\\pr33}}+C_{\substack{lu\\pr23}} \dfrac{V^*_{cs}}{V^*_{ts}} \dfrac{y_c}{y_t}\Bigg{]}\,.
\end{split}
\end{align}
We have neglected the tiny QED running below $\mu=\mu_\mathrm{EW}$. 
The above equation involves the right combination of WCs needed to explain a deficit of $R_{K^{(\ast)}}$, cf.~Sec.~\ref{ssec:lowenergy} and Table \ref{tab:wc-tree-fit}. 
Note that the mixed loop with a charm and top quark induces a non-negligible contribution, since the CKM factor $V_{cs}^*/V_{ts}^*$ partially compensates the $y_c/y_t$ suppression. This feature was first pointed out in Ref.~\cite{Becirevic:2017jtw}, which considered a concrete model, and further discussed in Ref.~\cite{Camargo-Molina:2018cwu}.

The most important constraint on this scenario arises at loop level, from the modification of the $Z$-boson couplings, as depicted in Fig.~\ref{fig:diagrams-Olu}. Working under the same approximations as above, we obtain the following contribution at $\mu=\mu_{\mathrm{EW}}$,
 
\begin{align}
\begin{split}
{\cal L}_{\text{SMEFT}}&\supset \dfrac{\log \left(\Lambda/m_t\right)}{16 \pi^2 \Lambda^2}\, 6 \, C_{\substack{lu \\ prvw}} [Y_u\,Y_u^\dagger]_{wv}\,\mathcal{O}_{\substack{Hl\\pr}}^{(1)}\,,\\
&\simeq \dfrac{\log \left(\Lambda/m_t\right)}{16 \pi^2 \Lambda^2}\, 6 \, y_t^2 \, C_{\substack{lu \\ pr33}} \,\mathcal{O}_{\substack{Hl\\pr}}^{(1)}\,,
\end{split}
\end{align}
where $\mathcal{O}_{Hl}^{(1)}$ is defined in Eq.~\eqref{eq:OHl1}. 
The only significant term arises from the top-quark loop. Recalling the discussion above Eq.~\eqref{eq:Z-pole-example}, we obtain from LEP data that
\begin{equation}
C_{\substack{lu \\ 2233}} \in (-0.76,0.36) \times \left(\dfrac{\Lambda}{1~\mathrm{TeV}}\right)^2\,,
\end{equation}
where we fixed $\Lambda =1$~TeV in the logarithm. On the other hand, the quark-flavour-violating WC appearing in Eq.~\eqref{eq:C9mC10-Olu} is not constrained by $Z$-pole observables.

\begin{figure*}[!t]
\centering
\includegraphics[width=0.45\textwidth]{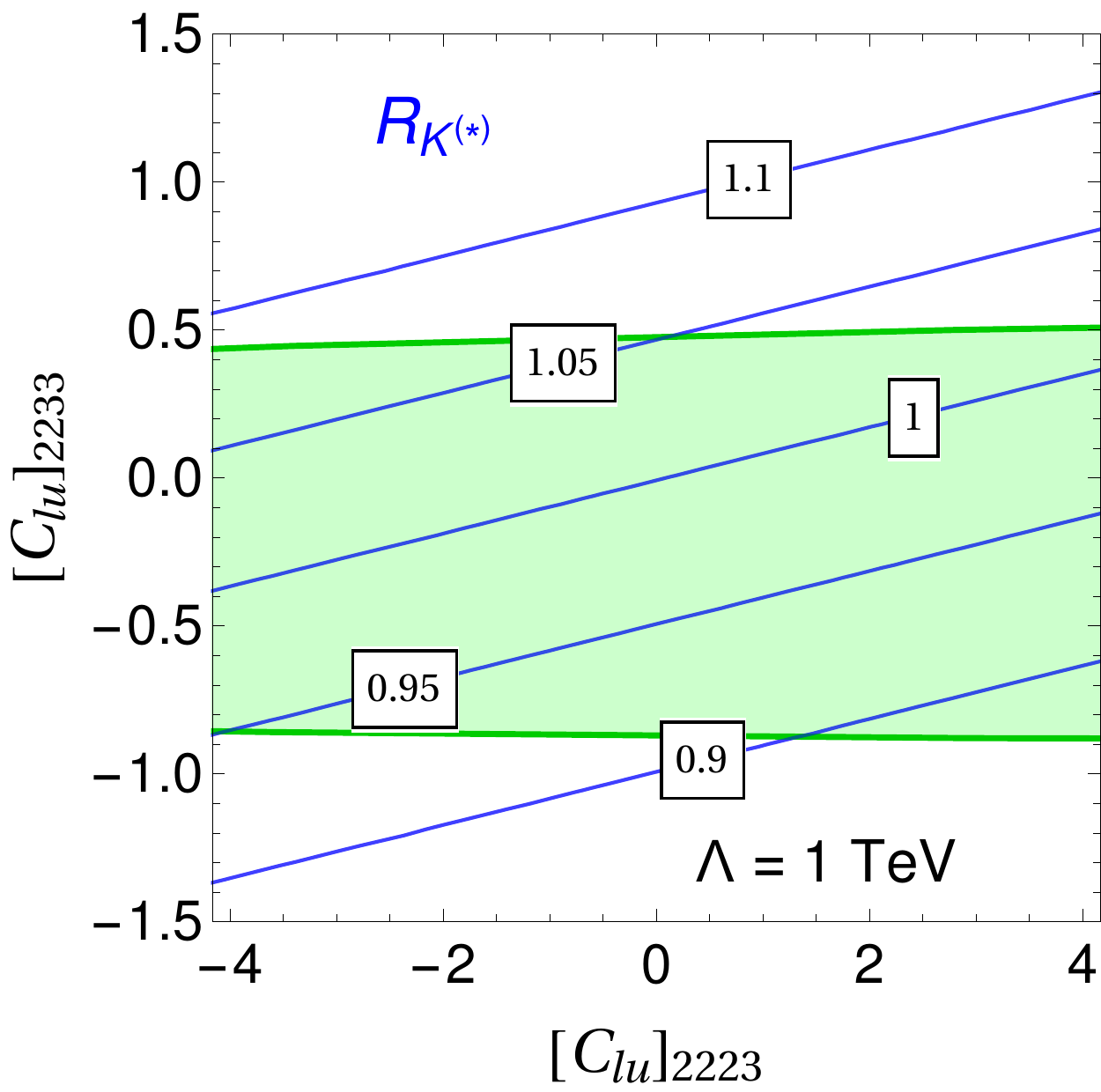}~\includegraphics[width=0.45\textwidth]{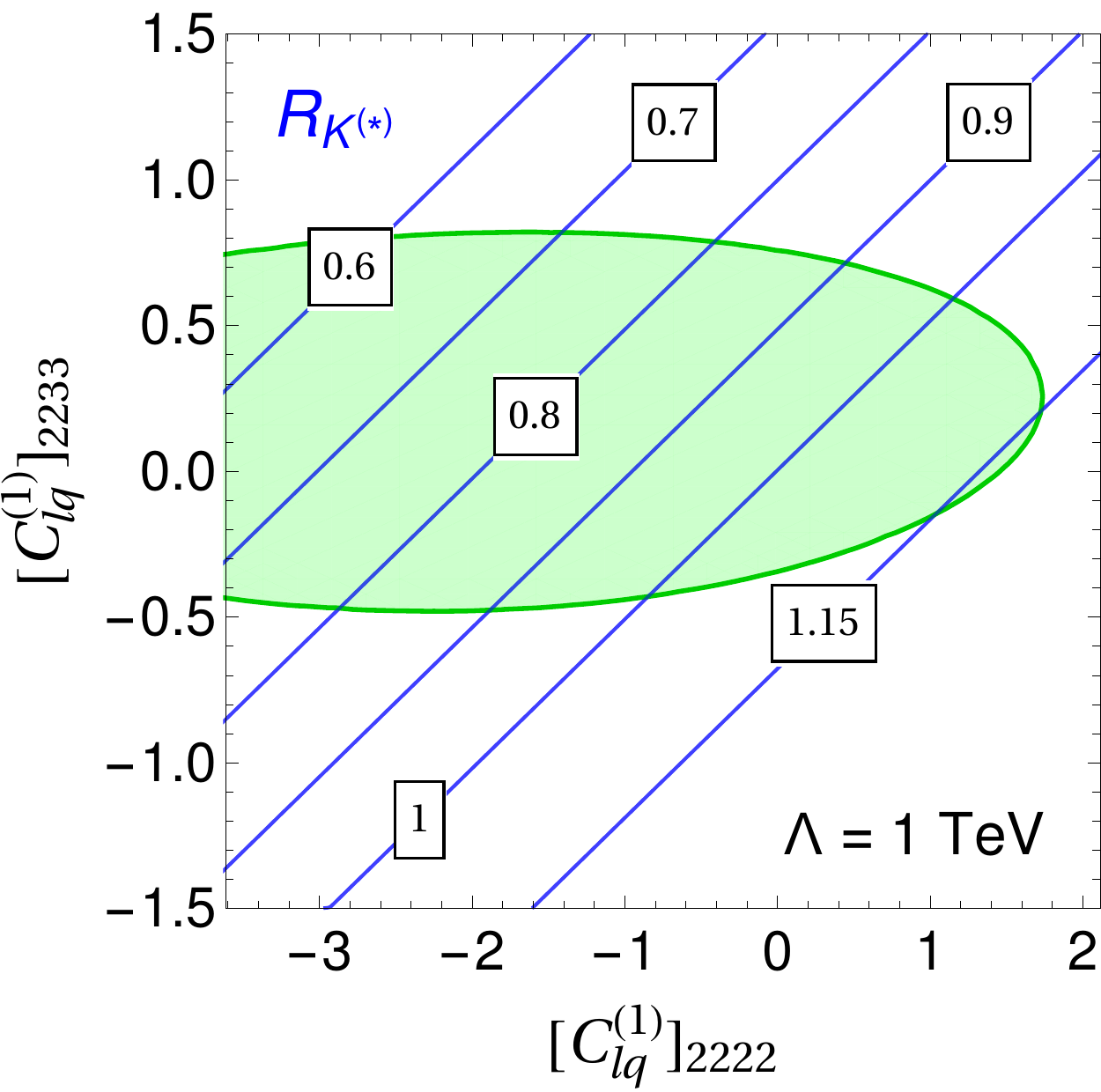}
\caption{\sl Predictions for $R_{K^{(\ast)}}$ in the central bin (blue lines) coming from the running of WCs $C_{lu}$ (left panel) and $C_{lq}^{(1)}$ (right panel) coupled to muons, taking $\Lambda = 1$~TeV. The green band corresponds to region allowed by the constraints listed in Sec.~\ref{ssec:constraints}. See text for details. }
\label{fig:predictions-Olu}
\end{figure*}

The constraints discussed above are combined in Fig.~\ref{fig:predictions-Olu} to show the valid range of WCs in the muon sector, and to predict the allowed contributions to $R_{K^{(\ast)}}$ in the central $q^2$ bin. From this plot, we see that $R_{K^{(\ast)}}$ has a strong dependence on the effective coefficient with the top quark, which, as discussed above, is tightly constrained by LEP. 
Conversely, it shows only a mild dependence on the quark-flavour-violating WC, which is poorly constrained by low-energy data.  
We find that $\mathcal{O}(1)$ couplings can produce a $\mathcal{O}(10\%)$ deficit in $R_{K^{(\ast)}}$, in agreement with the latest $R_K$ measurement by LHCb~\cite{Aaij:2019wad}. These conclusions have been obtained without considering LHC data. While high-$p_T$ dimuon tails can provide useful limits on this scenario, their precise assessment would require us to specify an ultraviolet completion, since LHC energies lie beyond the regime of validity of our EFT. 
We postpone this task to Sec.~\ref{sec:simplified}, where specific mediators are considered.

\subsection{\underline{$O_{lq}^{(1)} = (\bar{l} \gamma^\mu l)(\bar{q} \gamma_\mu q)$} \Big{[}and $\underline{O_{lq}^{(3)} = (\bar{l} \gamma^\mu \sigma^I l)(\bar{q} \gamma_\mu \sigma^I q)}$ \Big{]}} 
\label{ssec:Olq}

Another viable scenario that we point out here, for the first time, is the one with a purely left-handed operator, $\mathcal{O}_{lq}^{(1)}$, with a flavour structure that suppresses or forbids the tree-level contribution to $b\to s \ell\ell$. Such a flavour structure could be realised e.g.~by mediators with predominant couplings to top-quarks and muons.\footnote{See Ref.~\cite{Crivellin:2018yvo} for a related discussion where large couplings to third-generation of quarks and leptons induce a measurable LFU contribution to $b\to s\ell\ell$.} For sake of generality, we also consider the operator $\mathcal{O}_{lq}^{(3)}$, which is predicted together with $\mathcal{O}_{lq}^{(1)}$ in several models, cf.~Sec.~\ref{sec:simplified}.

\begin{figure}[b]
\centering
\includegraphics[width=0.5\textwidth]{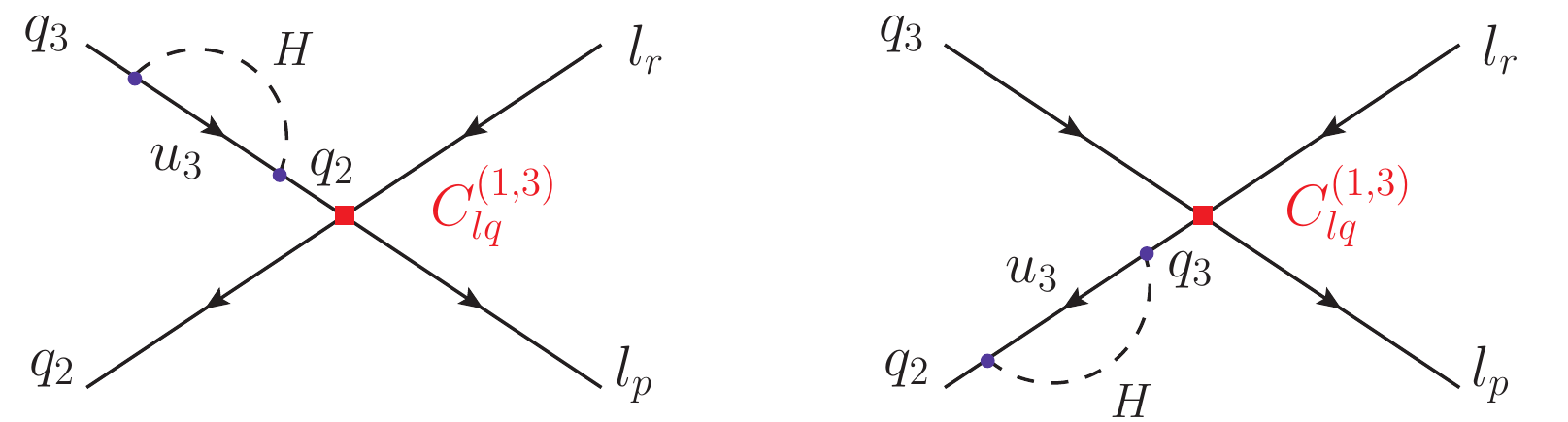}
\caption{\sl RGE-induced mixing of the operators $\mathcal{O}_{lq}^{(1,3)}$, with quark flavour indices $(22)$ and $(33)$, into the same operator with indices $(23)$, which contributes to $R_{K^{(\ast)}}$.}
\label{fig:diagrams-Olq}
\end{figure}

The RGE from $\mu=\Lambda$ down to $\mu_{\mathrm{EW}}$ modifies the WCs of the ${\cal O}_{lq}^{(1,3)}$ operators, as illustrated in Fig.~\ref{fig:diagrams-Olq}. The relevant Lagrangian at $\mu = \mu_\mathrm{EW}$ can then be written as

\begin{align}
\label{eq:lqRGE}
{\cal L}_{\text{SMEFT}} \supset &  \sum_{a=1,3}\frac{1}{\Lambda^2} \Bigg{\lbrace}
C^{(a)}_{\substack{lq \\ prst}} - \dfrac{1}{32 \pi^2}\log \frac{\Lambda}{m_t} \\
\times & \left[ (Y_u^\dagger Y_u)_{sv} C_{\substack{lq \\ prvt}}^{(a)} + C_{\substack{lq \\ prsv}}^{(a)} (Y_u^\dagger Y_u)_{vt} \right] \Bigg{\rbrace}
\,\mathcal{O}_{\substack{lq\\prst}}^{(a)}\,,\nonumber\end{align}
where the first term corresponds to the tree-level contribution and the others come from the one-loop RGEs. 
Besides these effects, it is crucial to account for the running of the down-quark Yukawa matrix, $Y_d$, which induces similar size effects in this specific scenario, as we now describe. 

We assume that $Y_d=\hat{Y}_d$ is diagonal at the scale $\Lambda$ and we will quantify the modification stemming from the SM Yukawa running to Eq.~\eqref{eq:lqRGE}.
This effect is described in the SM at one-loop by~\cite{Machacek:1983tz}
\begin{align}
\begin{split}
16 \pi^2 \dfrac{\mathrm{d}\,Y_d}{\mathrm{d}\log \mu} &\simeq \dfrac{3}{2} \left(Y_d Y_d^\dagger Y_d-Y_d Y_u^\dagger Y_u\right)\\
&+3\,\mathrm{Tr}\left[Y_u^\dagger Y_d+Y_d^\dagger Y_d\right] Y_d - 8 g_3^2 Y_d\,,
\end{split}
\end{align}

\noindent where the electroweak couplings and lepton Yukawas have been neglected. The running from $\mu=\Lambda$ to the electroweak scale induces an off-diagonal entry, namely
\begin{align}
\left(Y_d\right)_{32}\Big{\vert}_{\mu = m_t} \simeq \dfrac{3 \,V_{tb}^\ast V_{ts}\,y_b^\prime y_t^{\prime\,2}}{32\pi^2} \log \dfrac{\Lambda}{m_t}\,,
\end{align}
\noindent where the primed Yukawas are defined at $\Lambda$, and where we have kept only the dominant effects. Since we are interested in FCNC effects in the down sector, the matrix $Y_d$ should be rediagonalised at the electroweak scale. 
This is achieved by a redefinition of the quark doublets, which requires a change of flavour basis in Eq.~\eqref{eq:lqRGE}. Thus, the contribution of SMEFT operators with quark-flavour indices $22$ and $33$ to the WCs of the weak effective theory is
\begin{align}
\begin{split}
C_9^{pr} &= - C_{10}^{pr}
 \simeq  - \dfrac{m_t^2}{16\pi \alpha_{\mathrm{em}}\Lambda^2} 
\log \frac{\Lambda}{m_t} \left( \Delta^{pr}_{\text{mix}}+\Delta^{pr}_{\text{diag}} \right) \,,
\end{split}
\label{eq:C9C10-Olq1}
\end{align}
where the matching of Eq.~\eqref{eq:lqRGE} gives
\begin{align}
\begin{split}
\Delta^{pr}_{\text{mix}} = \left(C_{\substack{lq\\pr33}}^{(1)}+ C_{\substack{lq\\pr22}}^{(1)} + C_{\substack{lq\\pr33}}^{(3)} + C_{\substack{lq\\pr22}}^{(3)}\right) \,,
\end{split}
\end{align}
while the contribution which is induced by the SM Yukawa running and quark doublet redefinition at $\mu_{\mathrm{EW}}$ is
\begin{align}
\begin{split}
\Delta^{pr}_{\text{diag}} = 3 \,\left(C_{\substack{lq\\pr33}}^{(1)}- C_{\substack{lq\\pr22}}^{(1)} + C_{\substack{lq\\pr33}}^{(3)} - C_{\substack{lq\\pr22}}^{(3)}\right) \,.
\end{split}
\end{align}

\noindent We see that the two effects are of the same order, in fact the diagonalisation gives a larger contribution than the mixing.
This $Y_d$ running is also important for the other semi-leptonic operator containing quark doublets, $\mathcal{O}_{eq}$. We accounted for these effects in Table~\ref{tab:wc-pred-loop} by using the package \emph{Wilson}~\cite{Aebischer:2018bkb}, finding good agreement with the analytical expressions given above. 

Before quantifying their impact onto flavour data, it should be stressed that the misalignment between mass and flavour basis has been considered before as a way of relating flavour-conserving WCs, coupled only to the third generation of fermions, to flavour violation in the $b\to s\ell\ell$ transition, cf.~e.g.~Ref.~\cite{Glashow:2014iga}. Here, we estimate the irreducible misalignment in the quark sector stemming from SM RG running, which should be added on top of tree-level mixing angles in concrete scenarios.


We now turn to constraints on this scenario. The WC
$C_{lq}^{(3)}$ is bounded at tree-level by LFU tests in meson decays. The other crucial limit arises from $Z$-pole observables, cf.~Ref.~\cite{Feruglio:2016gvd}. These observables are affected at $\mu = \mu_{\mathrm{EW}}$ by the RGE contributions,
\begin{align}
\begin{split}
{\cal L}_{\text{SMEFT}}
\simeq  \dfrac{\log \left(\Lambda/m_t\right)}{16 \pi^2 \Lambda^2} \, 6 \,  y_t^2 |V_{tb}|^2
 \Bigg{[}C_{\substack{lq \\ pr33}}^{(1)}\mathcal{O}_{\substack{Hl\\pr}}^{(1)}-C_{\substack{lq \\ pr33}}^{(3)}\mathcal{O}_{\substack{Hl\\pr}}^{(3)}\Bigg{]}\,.
\end{split}
\end{align}
which are combined with other low-energy constraints to determine the allowed parameter space (green region) in the right panel of Fig.~\ref{fig:predictions-Olu}. The $R_{K^{(\ast)}}$ contours in the same plot show that this scenario can produce a deficit as large as $40\%$ for $\mathcal{O}(1)$ couplings.\footnote{Note that the combination $C_{lq}^{(1)}=C_{lq}^{(3)}$ can produce equally large effects for $R_{K^{(\ast)}}$, cf.~Table~\ref{tab:wc-pred-loop}. 
In particular, this linear combination mixes into $C_{Hl}^{(1)} - C_{Hl}^{(3)}$, which is weakly constrained by $Z$-pole data, cf. Eq. \eqref{eq:Z-pole-example-2}.}  
These contributions can be larger than the ones in the $\mathcal{O}_{lu}$ scenario, as can be seen by comparing the two panels in Fig.~\ref{fig:predictions-Olu}.

\subsection{Complementary observables}

Before discussing the matching of the above operators onto concrete models, we comment on other flavour observables that might be modified at loop level. 
First, we have explicitly checked that $\mathcal{B}(K\to \pi \nu \overline{\nu})$ and $\mathcal{B}(B \to K \nu \overline{\nu})$ will receive contributions smaller than $\mathcal{O}(10\%)$ compared to the SM predictions, from the same loops shown in Figs.~\ref{fig:diagrams-Olu} and \ref{fig:diagrams-Olq}.\footnote{See Ref.~\cite{Bordone:2017lsy} for other studies relating $R_{K^{(\ast)}}$ to $K\to \pi \nu \bar{\nu}$.} These values are smaller than the planned sensitivity of NA62~\cite{CortinaGil:2018fkc} and Belle-II~\cite{Kou:2018nap} experiments, thus are difficult to probe in the coming years.

Another potential probe of these scenarios is the muon $g-2$, which currently shows a $\approx 3.7 \sigma$ discrepancy with respect to the SM, $\Delta a_\mu=a_\mu^{\mathrm{exp}} - a_\mu^{\mathrm{SM}} = (2.74 \pm 0.73) \times 10^{-9}$~\cite{Bennett:2006fi,Jegerlehner:2009ry,Blum:2018mom}. 
The WCs identified above can generate contributions to $a_\mu$ at two-loop leading-log order. 
However, since $\mathcal{O}_{lq}^{(1,3)}$ and $\mathcal{O}_{lu}$ are chirality-conserving, this effect is further suppressed by $m_\mu^2$. 
Thus, given the bounds discussed in Sec. \ref{ssec:constraints}, only a negligible shift in $a_\mu$ is permitted.

\section{From EFT to single-mediator models}
\label{sec:simplified}

In this section we study minimal single-mediator models that can generate the viable effective scenarios identified in the previous section, namely ${\cal O}_{lq}^{(1)}$ or ${\cal O}_{lu}$.\footnote{For previous one-loop explanations of $R_{K^{(\ast)}}$ in the literature, see Refs.~\cite{Belanger:2015nma,Bauer:2015knc,Becirevic:2017jtw,Kamenik:2017tnu}.} 
We remain in the basis where $Y_d$ is diagonal at $\Lambda$, now identifying this scale as the mediator mass. 
For minimality, we restrict ourselves to (i) leptoquarks (LQs) with a single Yukawa coupling, or (ii) a neutral $Z'$ gauge boson with one coupling to quarks and one to leptons. We will match these mediators onto the SMEFT at tree-level, verifying our results with \cite{deBlas:2017xtg}, and compute the shift $\delta C_9^{ii} = - \delta C_{10}^{ii}$ at one-loop leading-log order. Although models with a single vector resonance (either a vector LQ or a $Z'$) are not UV-complete, a consistent completion can be built in several scenarios \cite{Barbieri:2015yvd,Kamenik:2017tnu}. 
We assume that the relevant phenomenology is determined to good accuracy by the mass and coupling(s) of a single state. 

On top of the various constraints discussed in the context of our EFT analysis, we apply additional bounds to the single-mediator scenarios, because
\begin{itemize}
\item The mediator can be directly produced at colliders;
\item The mediator couplings may induce additional WCs, besides the one needed to explain $R_{K^{(*)}}$, contributing to other low-energy flavour observables;
\item LHC dilepton searches at high $p_T$ are sensitive to the specific mediator propagator.
\end{itemize} 
Considering this ensemble of constraints, we find two scenarios which give a net pull against the SM larger than $3\sigma$. Following the notation of Ref.~\cite{Dorsner:2016wpm}, these are
\begin{itemize}
\item $S_3 \sim (\overline{3},3)_{1/3}$ scalar LQ coupled to $q_3l_2$;
\item $U_1^\mu \sim (3,1)_{2/3}$ vector LQ coupled to $q_3l_1$,
\end{itemize}
while the $Z'^\mu \sim (1,1)_0$ vector boson coupled to $l_2l_2$ and $u_2u_3$ is a marginally successful case.
We indicated the SM representation of the mediator in the form $(SU(3)_c, SU(2)_L )_{Y}$, and we listed only the couplings sufficient for a good fit.

In the following, we provide a detailed discussion of 
why these three cases above stand out.
We will also mention an additional viable scenario, namely a finite one-loop contribution induced by the $S_1 \sim (\overline{3},1)_{1/3}$ scalar LQ coupled to $q_3 l_1$.


\subsection{Mediators for ${\cal O}^{(1)}_{lq}$ \Big{[}and ${\cal O}^{(3)}_{lq}$\Big{]}}
\label{V.1}

\begin{figure*}
\begin{center}
\includegraphics[width=0.495\textwidth]{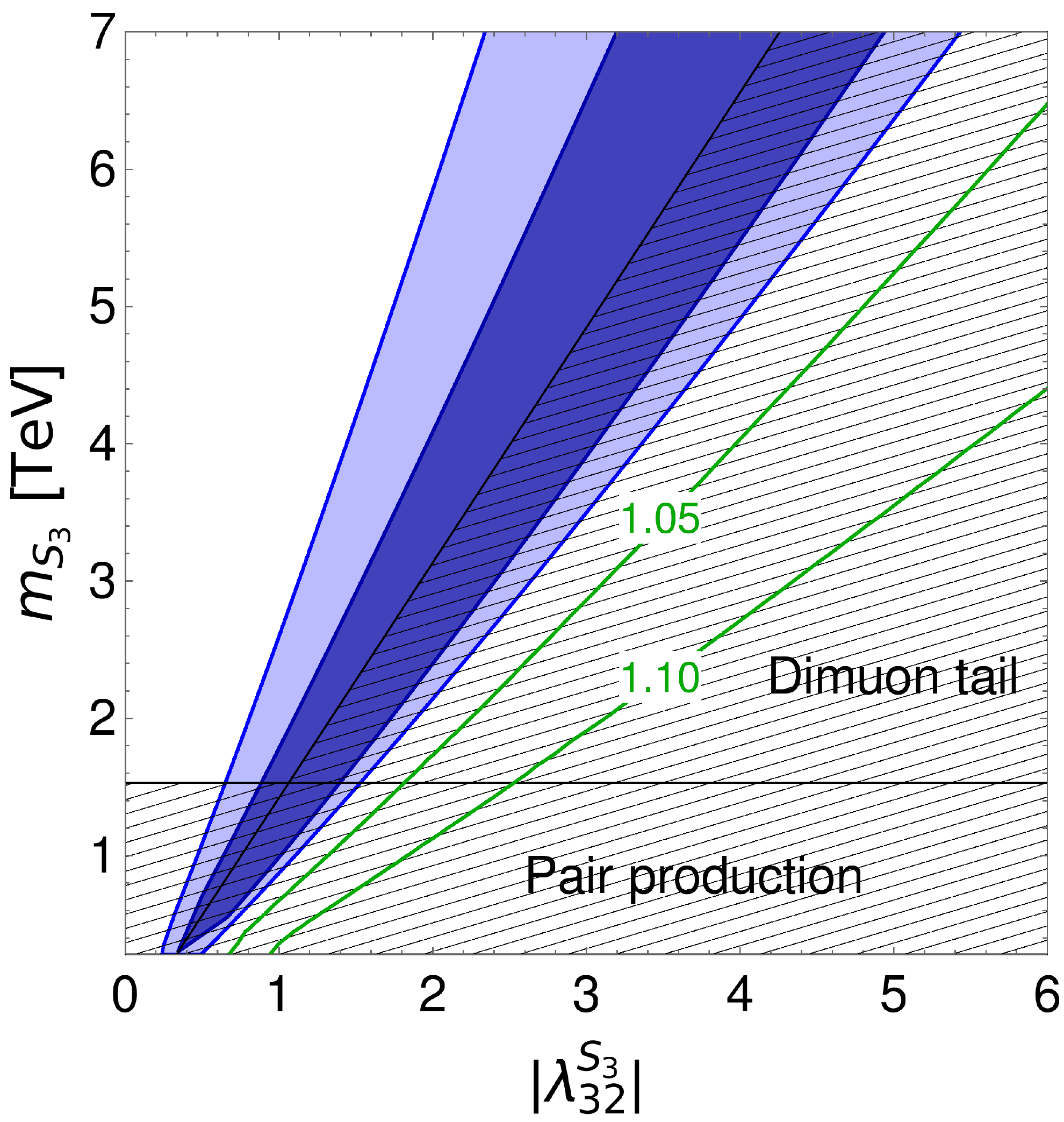}~\includegraphics[width=0.495\textwidth]{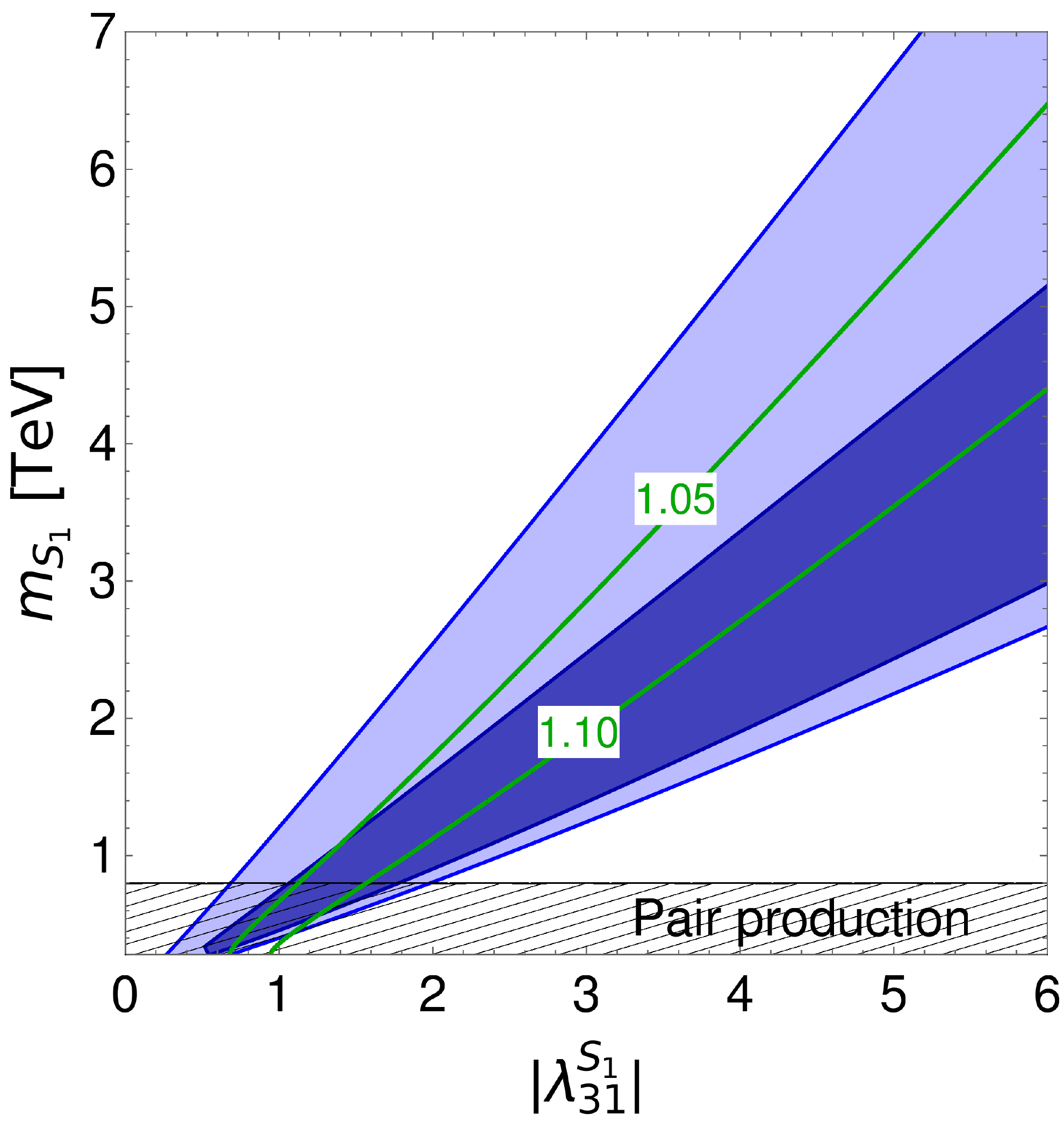}
\caption{Parameter space for scalar leptoquarks with a single coupling: $S_3$ with mass $m_{S_3}$ and coupling $\lambda^{S_3}_{32}$ (left panel), and $S_1$ with mass $m_{S_1}$ and coupling $\lambda^{S_1}_{31}$ (right panel). 
Dark (light) blue areas indicate the $1(2) \sigma$ preferred regions in our fit of the ensemble of flavour and electroweak precision observables described in subsection \ref{ssec:constraints}.  
Shaded regions are excluded by various collider constraints at 95\% C.L..
The green lines are contours for $R_K^\nu = R_{K^*}^\nu = 1.05, 1.1$.}
\label{s3Plot}
\end{center}
\end{figure*}

We start by discussing the scalar LQ, $S_3$. The relevant Lagrangian for our analysis is given by 
\begin{align}
\mathcal{L} & \supset  - m_{S_3}^2\, S_3^\dag S_3 + \left[ \lambda^{S_3}_{ij} \, \overline{q_{i}^c} (i\sigma_2\sigma^A) l_{j} S_3^A+ \mathrm{h.c.}\right]\,,
\label{S3}
\end{align}
where $\lambda^{S_3}_{ij}$ denotes the LQ Yukawa couplings. 
For a unique non-zero $\lambda^{S_3}_{ij}$, the  tree-level matching at $\mu=\Lambda$ gives the WCs 
\begin{equation}
\label{eq:matching-S3}
\frac{1}{\Lambda^2} C_{lq\atop{jj ii}}^{(1)} = \frac{3}{\Lambda^2} C_{lq\atop{jj ii}}^{(3)} = \frac{3 |\lambda^{S_3}_{ij}|^2}{4m_{S_3}^2} ~.
\end{equation}
Operator mixing then generates one-loop contributions to $b\to s$ transitions, inducing nonzero $C_9^{jj} - C_{10}^{jj}$, as explained in the previous section. We find a pull larger than $3\sigma$ with respect to the SM for a nonzero $\lambda^{S_3}_{32}$ coupling, i.e. with third-generation quarks running in the loop. 
The results are illustrated in the left panel of Fig.~\ref{s3Plot}, where we superimpose the result from our fit to flavour and electroweak precision observables with LHC constraints. 
These can be either limits from direct searches for pair-produced LQs or from the study of high-$p_T$ dimuon tails, which receive a $t$-channel LQ contribution. The $S_3$ with $\lambda^{S_3}_{32} \neq 0$ is constrained to $m_{S_3} \gtrsim 1400$ GeV at $95\%$ C.L by searches for the decay $S_3^{4/3} \to \mu^+ \bar{b}$~\cite{CMS:2018itt}. On the other hand, a reanalysis of the dimuon tail in Ref.~\cite{Angelescu:2018tyl} allows us to constrain a combination of $|\lambda_{32}^{S_3}|$ and $m_S$. 
From Fig.~\ref{s3Plot}, we see that LHC constraints probe an important fraction of the allowed parameter space, but this scenario remains a viable loop-level explanation of $R_{K^{(\ast)}}$.

The relevant interactions for the vector LQ, $U_1^\mu$, are
\begin{align}
\mathcal{L} & \supset  m_{U_1}^2 \,U_{1\mu}^\dagger U_1^\mu  + \left[\lambda^U_{i j} \, \overline{q_{i}} \gamma_\mu U^\mu_1 l_{j}+ \mathrm{h.c.}\right] ~.
\label{U1mu}
\end{align}
The tree-level matching generates 
\begin{equation}
\label{eq:clq13-u1}
\frac{1}{\Lambda^2} C_{lq\atop{jj ii}}^{(1)} = \frac{1}{\Lambda^2} C_{lq\atop{jj ii}}^{(3)} = - \frac{|\lambda^{U}_{ij}|^2}{2m_{U_1}^2} ~,
\end{equation}
where we obtain a different sign to Eq.~\eqref{eq:matching-S3}. Due to this sign difference, we find a pull larger than $3\sigma$ with respect to the SM in the scenario with $\lambda^U_{31}$, i.e. coupling to electrons rather than muons, unlike the $S_3$ case discussed above. 
This model can explain $R_{K^{(\ast)}}$ while remaining consistent with present LHC limits~\cite{CMS:2018bhq}. The parameter space is qualitatively similar to the $S_3$ case displayed in the left panel of Fig.~\ref{s3Plot}.\footnote{Since constraints from dilepton tails were not derived for electron couplings in Ref.~\cite{Angelescu:2018tyl}, we used the EFT bound from \cite{Greljo:2017vvb}, which is expected to hold up to {a $\mathcal{O}(1)$} factor. 
In this case, the allowed window is a little narrower than for the $S_3$.}

We remark that these minimal scenarios neatly avoid the most serious flavour bounds. 
Since $b \to s$ is generated at one loop, strongly-constrained $\Delta F = 2$ processes such as $K - \bar{K}$ mixing are generated at two loops, hence the bounds are easily satisfied by both models. 
The process $B \to K^{(*)} \nu \overline{\nu}$ is not induced by $U_1$ at one-loop leading-log order.  Moreover, the shift due to $S_3$ to $R_{K^{(\ast)}}^\nu \equiv \mathcal{B}(B\to K^{(\ast)}\nu\bar{\nu})/\mathcal{B}(B\to K^{(\ast)}\nu\bar{\nu})^\mathrm{SM}$ turns out to be very small and well below the experimental limits, $R_{K^*}^\nu < 2.7$ and $R_K^\nu < 3.9$~\cite{Buras:2014fpa,Grygier:2017tzo}, as shown by the green contour lines in Fig.~\ref{s3Plot}.

Let us now discuss a scenario in which the anomalies are explained by a one-loop finite LQ contribution, thus illustrating a limitation of our RGE analysis. Consider the $S_1 \sim (\overline{3},1)_{1/3}$ leptoquark with couplings only to fermion doublets,
\begin{equation}
\mathcal{L} \supset - m_{S_1}^2 S_1^\dag S_1 + \left[ \lambda^{S_1}_{ij} \overline{q_{i}^c} (i\sigma_2) l_j S_1  + \mathrm{h.c.} \right] ~.
\end{equation}
This does not contribute to $b \to s \ell_j \ell_j$ processes at tree-level, because it induces $C^{(1)}_{lq}= -C^{(3)}_{lq}$, and therefore $\delta C_9^{jj} = \delta C_{10}^{jj} = 0$. Nonetheless, as observed in Ref.~\cite{Bauer:2015knc}, this LQ gives a one-loop finite contribution to $C_9 - C_{10}$. For instance, by taking $\lambda^{S_1}_{31} \neq 0$, one obtains
\begin{equation}
\frac{1}{\Lambda^2} \left( C_{lq \atop{1123}}^{(1)} + C_{lq \atop{1123}}^{(3)} \right) = \frac{V_{ts}^* V_{tb} \, y_t^2 | \lambda^{S_1}_{31}|^2}{32 \pi^2 m_{S_1}^2}~.
\end{equation}
We verified that with the recently updated data summarised in Section \ref{sec:efts}, this scenario can explain the anomalies while obeying various constraints. 
These include the mild bound $m_{S_1} > 800$ GeV \cite{Sirunyan:2018kzh} from LHC searches for pair-produced $S_1$ decaying into a $b \overline{b} \nu \overline{\nu}$ final state.  Since there is no $e^+ e^- \to b \overline{b}$ at tree-level, the LEP (LHC) bounds from this (the reverse) process are negligible. 
Moreover, we did not find relevant constraints on the interactions $t\bar t\nu\bar\nu$ or $t\bar t e^+e^-$.
This scenario provides a pull larger than $3\sigma$ with respect to the SM. The best-fit region is shown in the right panel of Fig.~\ref{s3Plot}. 
The model induces only a small shift in $R_{K^{(*)}}^\nu $, as shown in the figure.

For completeness, we remark that $C_{lq}^{(1)}$ can also be generated at tree-level by the exchange of the vector LQ, $U_3 \sim (3,3)_{2/3}$, with a single coupling, or by a $Z'$ coupled to quark and lepton doublets.  The former is constrained by corrections to $Z$-couplings and gives a pull of at most $ 2.3\sigma$ against the SM. 
The latter case, in which the $Z'$ couples to one flavour of leptons and one of quarks, does not give a big pull against the SM due to LEP and LHC bounds on contact interactions as outlined in Section \ref{ssec:constraints}. As emphasised previously, the LHC bounds should be treated with caution as they are generally outside the EFT regime of validity. 
However, for $s$-channel processes mediated by a $Z'$ they provide a conservative bound (see e.g. \cite{Greljo:2017vvb}), so can be used to test the model's validity.


%

\subsection{Mediators for ${\cal O}_{lu}$ }

Apart from several flavour components of $C_{lq}^{(1)}$, the other operator that can accommodate the anomalies at one loop, identified in Section \ref{sec:viable}, is $\mathcal{O}_{lu\atop{2223}}$. 
This operator can be generated by a $Z'$ model with interactions
\begin{align}
\mathcal{L} &\supset  \frac{m_{Z'}^2}{2}  Z'_\mu Z'^\mu  - \left[ g^l_{ii} Z'_\mu \overline{l_{i}} \gamma^\mu l_{i} 
+ g_{jk}^u Z'_\mu \overline{u_{j}} \gamma^\mu u_{k} +\mathrm{h.c}.\right] ~,
\end{align}
by taking $g^l_{\mu\mu}, g_{ct}^u \neq 0$. 
Thus, at tree-level we generate
\begin{equation}
\frac{1}{\Lambda^2}  C_{lu\atop{2223}}= 
 -\frac{g^l_{\mu\mu} \,g_{ct}^u}{m_{Z'}^2}\,,\qquad
\frac{1}{\Lambda^2}  C_{ll\atop{2222}}= 
 -\frac{(g^l_{\mu\mu})^2}{ 2m_{Z'}^2}\,.
\end{equation}

We open a parenthesis on the choice of non-zero couplings for the mediators. In this paper we do not investigate the non-trivial theory of flavour needed to induce only the desired couplings: flavour symmetries can generally be engineered for this purpose. 
In the case of a gauge-boson mediator, there is the additional issue of building an ultraviolet-complete gauge model, in which that specific gauge boson is the lightest new particle. 
It is instructive  to sketch a toy model that may lead to a light $Z'_\nu$ coupled to $\overline{c}\gamma^\nu t$ and $\overline{l_{2}}\gamma^\nu l_{2}$ only. To have an off-diagonal coupling only
(in the up-quark singlet sector), one needs to introduce a non-abelian gauge symmetry, minimally $SU(2)'$,
and to split the three gauge boson masses so that the lightest is identified with $Z'_\nu\equiv Z'^{1}_\nu$. This can be achieved by introducing a complex scalar $\phi\sim 2_{SU(2)'}$ and a real scalar $\Delta_A\sim 3_{SU(2)'}$,
coupled as $\rho[\phi^T(i\sigma_2)\Delta_A\sigma_A \phi + \mathrm{h.c.}]$, with $\rho$ a real mass parameter. While the vev of $\phi$ provides an equal mass to the three gauge bosons, the triplet vev turns out to align in the $\Delta_1$ direction,
and one can check that this contributes to the masses of $Z'^{2,3}_\nu$ only, making them parametrically heavier.
Now, any fermion $\psi \equiv (\psi_1 ~ \psi_2)^T \sim 2_{SU(2)^\prime}$ couples to $Z'^1_\nu$ off-diagonally, $g' Z'^1_\nu (\overline{\psi_1}\gamma^\nu \psi_2 + \overline{\psi_2}\gamma^\nu \psi_1)$. 
For the quark sector, one can identify $\psi$ with a vector-like up-quark singlet, $U$, and arrange for $c$ ($t$) to mix with $U_{1}$ ($U_{2}$) only via the vev of $\phi$.
For the lepton sector, the appropriate $\psi$ is a vector-like lepton doublet, $L$, with $l_{2}$ mixing with both $SU(2)^\prime$ components $L_{1}$ and $L_{2}$. These mixings can be arranged
by an appropriate flavour symmetry and provide the desired pattern of $Z'$ couplings. 
While such a UV completion is certainly not unique, it demonstrates that a $Z'$ with the required couplings can be the lightest new physics state.

\begin{figure}[tb!]
\begin{center}
\includegraphics[width=0.495\textwidth]{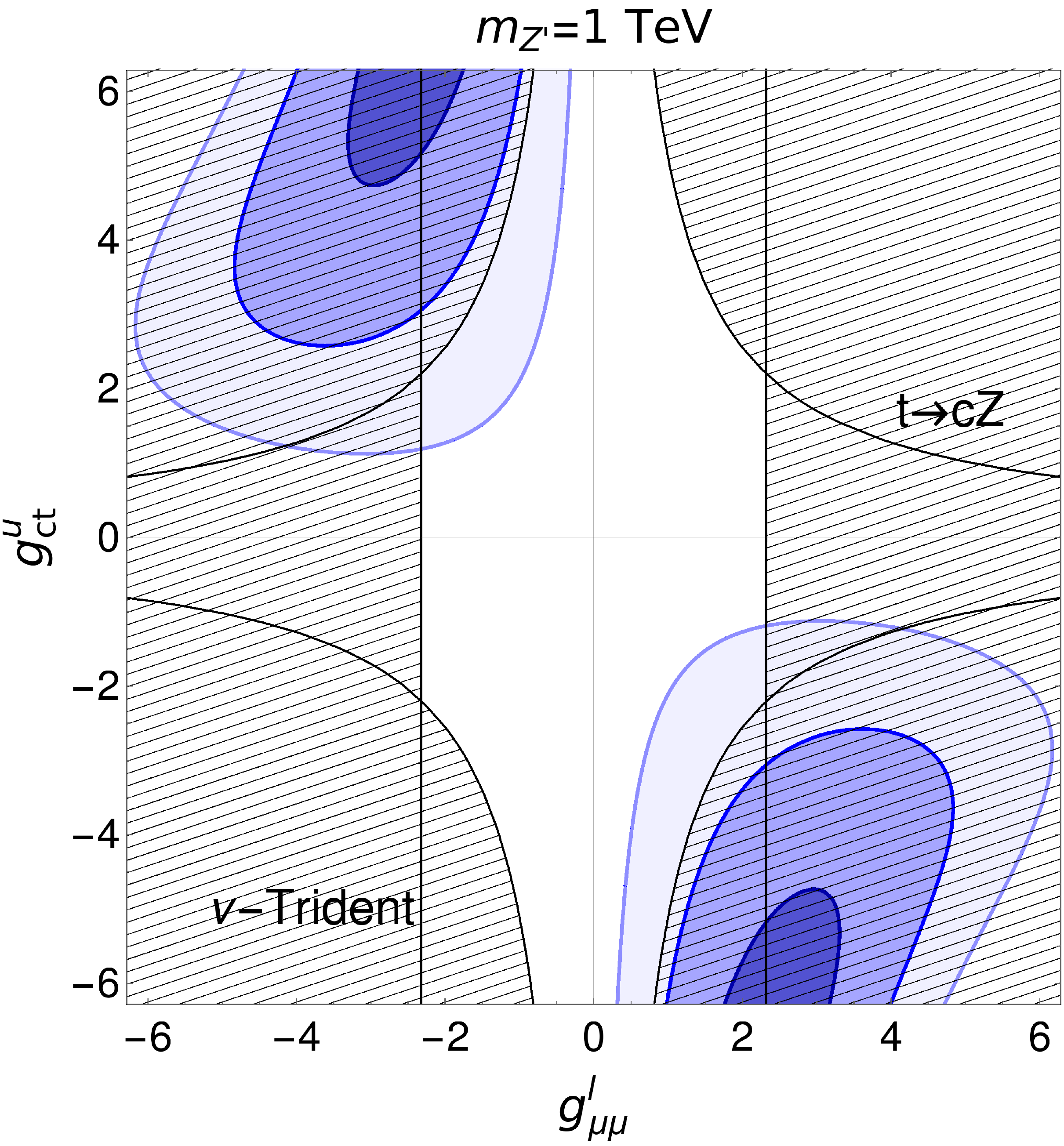}
\caption{Vector boson $Z'_\mu$ with $m_{Z'}=1$ TeV and couplings $g^l_{\mu\mu}$ and $g^u_{ct}$.
Dark, normal and light blue are the $1,2,3 \sigma$ preferred regions in our fit of flavour and electroweak observables. 
Shaded regions are excluded by other constraints. }
\label{zctplot}
\end{center}
\end{figure}

Let us discuss the experimental bounds on such model.
The main constraint on  $g^l_{\mu\mu}$ stems from the $\nu_\mu$ trident process, 
$ \nu_\mu + N \to \nu_\mu \mu^+ \mu^- + N$ \cite{Altmannshofer:2019zhy}. Using $\sigma^{\text{CHARM-II}}/\sigma^{\text{SM}}=1.58\pm 0.64$ from Ref.~\cite{Geiregat:1990gz} and $\sigma^{\text{CCFR}}/\sigma^{\text{SM}}=0.82\pm 0.28$ from Ref.~\cite{Mishra:1991bv} as experimental input, and the recent calculation of the $\nu_\mu$ trident cross-section in~\cite{Altmannshofer:2019zhy} as theoretical input, where subleading nucleus effects are included, we obtain
$|g^l_{\mu\mu}|\leq 2.3\, m_{Z'}/\text{TeV}$ at $2\sigma$. 
Furthermore, the upper limit on $t \to c Z [\to \ell^+ \ell^-]$ decays discussed in Sec. \ref{ssec:constraints}, $\mathcal{B}(t \to c Z) < 2.4 \times 10^{-4}$ at $95\%$ C.L. 
\cite{Aaboud:2018nyl}, can be reinterpreted as a search for $t \to c \mu^+ \mu^-$ decays mediated by a virtual $Z'$. 
This places an upper bound on $|g^l_{\mu\mu} g_{ct}^u| / m_{Z'}^2$. 
Taking the $Z'$ propagator to be $1/m_{Z'}^2$ is a good approximation, since the experiment makes a cut on the dimuon-pair invariant mass, 
$m_{\ell \ell} \in [m_Z - 15 \text{ GeV},m_Z + 15 \text{ GeV}] \ll m_{Z'}$. 
This bound cuts into the preferred parameter space, see Fig.~\ref{zctplot}. 
The 1 and 2$\sigma$ best-fit regions are excluded, thus this minimal scenario provides only a modest improvement ($\approx 2.4 \sigma$) over the SM with regards to the $b \to s$ anomalies.

Alternatively, the WC $C_{lu}$ could be generated by a scalar LQ, $R_2 \sim (3,2)_{7/6}$, or a vector LQ, $\tilde{V}_2 \sim (\overline{3},2)_{-1/6}$, with interactions
\begin{align}
\mathcal{L} \supset \lambda^R_{ij} \overline{u_{i}} R_2 l_{j} + \lambda^V_{ij} \tilde{V}_2^\mu \overline{u^c_{i}} \gamma_\mu l_{j} ~,
\end{align}
with $\lambda^R_{i\mu} \neq 0$ or $\lambda^V_{i\mu} \neq 0$, for $i = c,t$. 
The former was proposed as a loop solution in Ref. \cite{Becirevic:2017jtw}. While it remains possible with two (or more) couplings, we confirm that with only a single coupling it does not give a large pull against the SM due to a combination of $Z$-pole bounds and LHC constraints, cf. \cite{Camargo-Molina:2018cwu,Angelescu:2018tyl}. 
The $\tilde{V}_2$ scenario has not to our knowledge been considered in the literature, and we found that the combination of $Z$-pole and LHC bounds also rules out this case.

\section{Conclusion}
\label{sec:conclusion}

The current ensemble of $b \to s \ell \ell$ anomalies constitutes one of the most statistically significant departures from the SM in flavour data. 
In this article, we have comprehensively classified new physics explanations in the language of the SMEFT. 
After reviewing the tree-level solutions in Section \ref{sec:efts}, we performed a thorough analysis of the possible contributions at one-loop leading-log order in Section \ref{sec:loop-level}. 
We extended previous analyses by inspecting all possible WCs, and imposing a broader range of constraints, including bounds from $Z$-pole observables, LFU in meson decays, and collider bounds on contact interactions. 
In total, we found just a few individual WCs that provide a successful fit of the data, as summarised in Table \ref{tab:wc-pred-loop}. 
Apart from the $C_{lu}$ scenario, previously pointed out in the literature, we showed for the first time that $C_{lq}^{(1)}$ or $C_{lq}^{(1)}=C_{lq}^{(3)}$, with flavour-conserving couplings to quarks, can also explain the anomalies at loop level. 
The working scenarios were discussed in detail in Section \ref{sec:viable}, carefully including the running of the down-quark Yukawa, $Y_d$, between the new physics and the electroweak scale, which we found to be qualitatively important. 
We further demonstrated that the associated shifts in $\mathcal{B} (K \to \pi \nu \overline{\nu})$ and $\mathcal{B}(B \to K \nu \overline{\nu})$ are much smaller than their experimental sensitivities.

We exploited the working EFT scenarios to construct minimal UV-complete models in Section \ref{sec:simplified}. 
We considered models involving a single LQ ($Z’$) with only one (two) coupling(s) to SM fermions of definite flavour. 
Such minimal scenarios had not previously been considered in the literature, yet we demonstrated that three LQ scenarios are able to explain the $b \to s\ell\ell$ anomalies while conforming to both EFT and model-specific constraints. One $Z'$ scenario proved to be only marginally successful after we accounted for all constraints. 
The favoured parameter space is shown in Fig.~\ref{s3Plot} for the two scalar LQ models and in Fig.~\ref{zctplot} for for the $Z'$ model. 
This exercise highlights the usefulness of our EFT results for model-building.

A limitation of our analysis is that we do not account for finite one-loop contributions. One such case is provided by the $S_1$ LQ, as discussed in Section \ref{V.1}.
Other such cases cannot be excluded, but they
have to contend with the wide range of constraints which we outlined, and they are likely marginal.

The paucity of loop-level solutions which evade all bounds -- both in the EFT and the single-mediator analyses 
-- shows the difficulty in explaining the $b\to s\ell\ell$ anomalies
with TeV-scale new physics.
If the anomalies persist, we have shown that only very specific directions in the EFT parameter space are viable, and only very restricted model-building avenues can be taken. 
There is a significant chance of confirming or disproving these possibilities with the expected experimental progress in the near future.

\

\section*{Acknowledgments}

We thank S.~Davidson, N.~Kosnik and P.~Paradisi for discussions, and P.~Stangl, D.~Straub and J.~Virto for clarification regarding the packages \emph{flavio} and \emph{DsixTools}. O.S.~thanks the Universitat de Barcelona for the kind hospitality. This project has received support by the European Union's Horizon 2020 research and innovation programme under the Marie Sklodowska-Curie grant agreement N$^\circ$~674896 (ITN Elusives) and 690575 (RISE InvisiblePlus). 
M.F. is also supported by the OCEVU
Labex (ANR-11-LABX-0060) and the A*MIDEX project (ANR-11-IDEX-0001-02) funded by
the “Investissements d’Avenir” French government program managed by the ANR.
F.M. is supported by MINECO grant FPA2016-76005-C2-1-P, by Maria de Maetzu program grant MDM-2014-0367 of ICCUB and 2017 SGR 929.

\appendix

\section{Notation and conventions}
\label{app:conventions}

We consider the same notation of Ref.~\cite{Jenkins:2013zja,Jenkins:2013wua,Alonso:2013hga} for the operators in the Warsaw basis, except for the notation replacement $\mathcal{O}_{qe} \to \mathcal{O}_{eq}$, which ensures that lepton flavour indices come before quark flavour indices in all operators. Quark and lepton doublets are denoted by $q$ and $l$, while up and down quarks and lepton singlets are denoted by $u$, $d$ and $e$, respectively. Our convention for the covariant derivative is given by
\begin{equation}
D_\mu=\partial_\mu + i g_1 \, Y B_\mu + i g_2\, \tau^I W^I_\mu + i g_3\, T^A G_\mu^A \,,
\end{equation}
where $T^A=\lambda^A/2$ are the $SU(3)_c$ generators, $\tau^I = \sigma^I/2$ are the $SU(2)_L$ generators and $Y$ denotes the hypercharge. The Yukawa couplings are defined in flavour basis as
\begin{equation}
\mathcal{L}_{\mathrm{yuk}} = - H^\dagger\, \bar{d}\, Y_d\, q - \widetilde{H}^\dagger \,\bar{u}\, Y_u\, q - H^\dagger \, \bar{e}\, Y_e\, l+\mathrm{h.c.}\,,
\end{equation}

\noindent where flavour indices have been omitted. 
We work in the basis where $Y_\ell = \widehat{Y}_\ell$ and $Y_d = \widehat{Y}_d$ are diagonal matrices, while 
$Y_u = \widehat{Y}_u V$ depends on the CKM matrix, $V \equiv V_{\mathrm{CKM}}$.








\end{document}